\documentclass[aps,prb,groupedaddress,showpacs,superscriptaddress,amssymb,amsmath,longbibliography,]{revtex4-1}

\usepackage{physics}

\usepackage{amsmath}
\usepackage{mathptmx,bm} 
\usepackage{graphicx}
\usepackage{dcolumn}
\usepackage{tabularx}
\usepackage{epsf}
\usepackage{color}

\usepackage{hyperref}
\hypersetup{breaklinks,colorlinks,linkcolor=blue,citecolor=blue,urlcolor=blue}

\usepackage{verbatim}
\usepackage[english]{babel}
\newcommand{\bea}{\begin{eqnarray}}
\newcommand{\eea}{\end{eqnarray}}
\usepackage{amsfonts}

\begin{document}


\title{Strong coupling effects in quantum thermal transport with the reaction coordinate method}

\author{Nicholas Anto-Sztrikacs}
\affiliation{Department of Physics, 60 Saint George St., University of Toronto, Toronto, Ontario, Canada M5S 1A7}



\author{Dvira Segal}
\affiliation{Chemical Physics Theory Group, Department of Chemistry and Centre for Quantum Information and Quantum Control,
University of Toronto, 80 Saint George St., Toronto, Ontario, M5S 3H6, Canada}
\affiliation{Department of Physics, 60 Saint George St., University of Toronto, Toronto, Ontario, Canada M5S 1A7}
\email{dvira.segal@utoronto.ca}

\date{\today}

\begin{abstract}
We present a semi-analytical approach for studying quantum thermal energy transport beyond the weak
system-bath coupling regime.
Our treatment, which results in a renormalized, effective Hamiltonian model 
is based on the reaction coordinate method.
In our technique, applied to the nonequilibrium spin-boson model, a collective coordinate is extracted
from each environment and added into the system to construct an enlarged system.
After performing additional Hamiltonian's truncation and transformation,
we attain an effective two-level system with renormalized parameters, 
which is  weakly coupled to its environments, thus can be simulated
using a perturbative Markovian quantum master equation approach.
We compare the heat current characteristics in our method to other techniques,
and demonstrate that we properly capture strong system-bath signatures 
such as the turnover behavior of the heat current as a function of system-bath coupling strength. 
We further investigate the thermal diode effect and
demonstrate that strong couplings moderately improve the rectification ratio relative to the weak coupling limit. 
The effective Hamiltonian method that we developed here 
offers fundamental insight into the strong coupling behavior, and is computationally economic.
Applications of the method towards studying quantum thermal machines are anticipated. 
\end{abstract}

\maketitle

\section{Introduction}

Nonequilibrium dissipative quantum systems play a central role in quantum technologies.
However, simulating the dynamics and transport properties of open quantum systems
is challenging given the important role of the surrounding environment.
Perturbative treatments in the language of Green's functions \cite{diventra} or quantum master 
equations (QMEs) \cite{Breuer,Nitzan,Strunz20} have proved fruitful, bringing much insights. 
QME methods are perturbative in the coupling energy of the system to its surroundings,
further typically assuming a Markovian-memoryless system's dynamics.
Particularly for second-order QMEs, the rate constants responsible for 
transitions between quantum states are linear in the system-bath
interaction energy (quantified e.g. by the reorganization energy).
Nevertheless, many open-system processes fall outside the regime of validity of 
perturbation techniques, ranging from biological charge and exciton transport processes
\cite{Thorwart,Coker} to quantum devices \cite{Nori}. 

In contrast to the linear-monotonous nature of the weak coupling regime,
strong system-bath couplings can enact rich functional behavior relying on involved phenomena
such as time nonlocal effects, cooperative interactions between thermal baths,
and the buildup of correlations between a system and its surroundings.
Significant efforts are currently directed to understand---and harness---the nontrivial 
outcome of strong system-bath couplings on the performance of quantum heat machines;
a small sample of such studies include \cite{Strasberg_2016,Hava18,Newman,Goold,Misha,Wiedmann2020,Motz2018}.

The nonequilibrium spin-boson (SB) model, with a two-level system (spin) coupled to two independent 
harmonic-oscillator (HO) baths was suggested for exploring the fundamentals of thermal energy 
transport in anharmonic nanojunctions \cite{PRL05,QME06}.
Remarkably, the model was recently realized in superconducting quantum circuits
demonstrating a thermal diode effect \cite{PekolaE}.
The heat current in the SB model displays a turnover behavior at strong coupling:
While in the weak coupling limit the steady state heat current grows 
monotonically with the system-bath interaction energy,
at strong coupling the current is suppressed \cite{ThossMCTDH,NIBA11,Nazim14,NIBA14,ARPC,Aurell2020}.

Numerous techniques were developed to investigate 
the thermal transport characteristics of the nonequilibrium SB model,
specifically the turnover behavior mentioned above,
by going beyond the weak-coupling Born-Markov Redfield (BMR) equation.
For example, the non-interacting blip approximation (NIBA), which is perturbative 
in the tunneling splitting (nonadiabatic parameter) 
can properly describe strong system-bath coupling effects for Ohmic baths \cite{PRL05,QME06,NIBA11,NIBA14,Aurell2019}.
To interpolate between the weak-coupling Born-Markov Redfield equation and the strong-coupling NIBA
for general spectral functions, a nonequilibrium polaron-transformed Redfield equation 
has been developed in Refs. \cite{Cao1,Cao2}. Keldysh nonequilibrium Green's function (NEGF) methods 
were furthermore developed for the SB model in Refs. \cite{ThossNEGF,WuNEGF,NJP17,PT-NEGF17}. 

Beyond perturbative methods, numerically-exact techniques, originally developed 
to study the dissipative dynamics of the spin-boson model, were extended to investigate the behavior of 
the heat current in steady state. This includes simulations with
the multi-layer multi-configuration Hartree approach \cite{ThossMCTDH,MCTDHH}, 
iterative influence functional path integral techniques \cite{Segal13,Nazim14,PI}, 
quantum Monte Carlo \cite{Saito1,Saito2} and the hierarchical equation of motion 
\cite{HEOM1,HEOM2,HEOM3,HEOM4}. Other treatments involve semiclassical \cite{Kelly}, mixed quantum-classical approximations \cite{mixed1,mixed2,mixed3}, 
and chain mapping approaches \cite{Plenio1,Plenio2,Plenio3,Poletti2020}. 

The reaction coordinate (RC) method offers an alternative, balanced approach between low-order perturbation theory treatments and brute-force numerical methods.
In this semi-analytic method, a collective coordinate of the bath, termed reaction coordinate,
is identified and included into the quantum system.
This step, or mapping, is exact. 
Since the RC incorporates system-bath interaction energies,
the dynamics of the enlarged, ``super-system"
captures effects beyond weak coupling and beyond the Markovian approximation
even {\it when evolved approximately using a second-order Markovian QME} 
\cite{PlenioProof,Francesco20}.
Obviously, the principle of the RC method can be generalized with 
several discrete modes extracted from the baths and included either as part of the system 
 or as a primary non Markovian environment, further coupled to a secondary Markovian bath
\cite{Gauger15, Eisfeld15, Nori19,PlenioProof, Francesco20, Plenio1, Plenio2, Plenio3, Poletti2020}.

Apart from its computational simplicity, the reaction-coordinate quantum master equation (RC-QME) procedure
can bring insights on the system's dynamics and the evolution of system-environment 
correlations \cite{NazirPRA14,Nazir16}. 
The RC-QME method has been applied onto both bosonic and fermionic systems  
\cite{GernotF,GernotF2,Nazir18,GalperinF}.
Furthermore, recent studies with this method examined the role of strong coupling effects in the 
performance of autonomous \cite{Strasberg_2016} and four-stroke \cite{Nazir18} thermal machines.

In this paper we develop, test, and benchmark a RC quantum master equation method for nonequilibrium quantum 
heat transport problems.
Beyond numerical simulations, our objective is to gain insight on the role of 
strong system-bath couplings in the operation of thermal devices.

The turnover behavior of the heat current with system-bath coupling energy has been 
demonstrated with several methods, interpreted using a picture analogous to polaron transformation 
\cite{PRL05,QME06,NIBA11,NIBA14,Cao1,Cao2}.
Here, we explain the turnover behavior of the heat current with an alternative picture:
Employing the RC framework, we show that with a series of transformations and truncation 
we can convert the original-standard spin-boson model with strong system-bath couplings
into an effective spin-boson model (EFF-SB). In the new model, the spin splitting 
is renormalized (suppressed) with coupling energy. Other parameters that are renormalized or transformed are
the coupling parameters of the spin to the heat baths and the reservoirs' spectral density functions.
The new, effective SB model captures nontrivial-nonlinear effects---such as the suppression of the heat current at 
strong coupling---even at the level of a Markovian second-order QME.
Overall, our approach handles the nontrivial strong coupling regime 
with minimal cost.

Our work addresses the following questions regarding heat transport in nanojunctions:
(i) What is the origin of the heat current suppression at strong coupling? 
(ii) How does the current scale with different system's parameters?
(iii) What is the extent of the thermal rectification (diode) behavior? 
Does strong coupling assist the rectification effect?

The paper is organized as follows.
In Section \ref{sec-Model}, we describe the standard spin-boson model. 
We further describe the RC procedure, which takes us to an 
effective spin-boson model with renormalized parameters.
We describe the RC-QME method in Sec. \ref{sec-Redfield}.
Simulations are presented in Sec. \ref{sec-simulA} where we focus on gaining insights
on the role of strong coupling in transport and on benchmarking the RC-QME technique against other
methods.
In Sec. \ref{sec-simulB}, we present additional simulations of the asymmetric spin-boson junction,
and analyze its operation as a thermal diode in the strong coupling regime.
We conclude and discuss future directions in Section \ref{sec-sum}.

\begin{figure*}[htpb]
\centering
\includegraphics[scale=0.4]{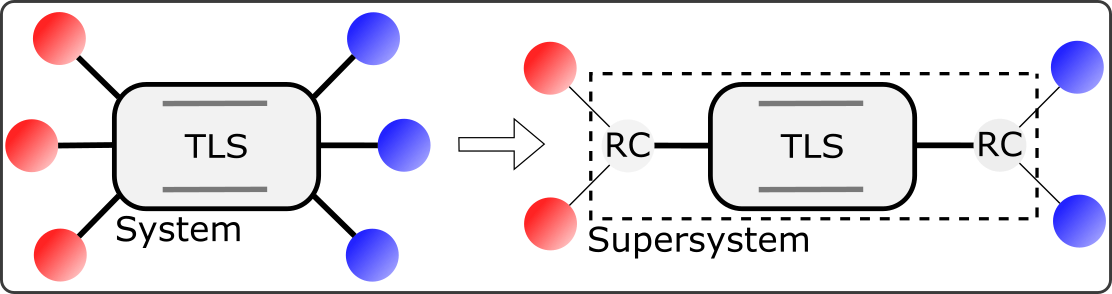}  
\caption{A two-level system interacting with two bosonic reservoirs held at different 
temperatures is mapped to an enlarged supersystem comprising the original two-level system 
coupled to two collective modes (one for each reservoir), which are 
in turn coupled to residual heat baths.}
\label{fig:RC_TLS}
\end{figure*}


\section{Model: Derivation of the effective spin-boson Hamiltonian}
\label{sec-Model}

We begin with the nonequilibrium spin-boson model allowing for strong coupling between 
the quantum system (spin) and the baths, 
which are characterized by peaked, Brownian oscillator spectral functions.
After a series of transformations and truncations, 
the model is reduced into an effective SB model,
which now can be assumed to be weakly coupled to the
baths---described by modified spectral functions.
The energy spacing of the effective spin and its coupling parameters to the baths 
are renormalized, thus providing understanding on the role of strong couplings in thermal transport.

\subsection{The standard SB model}
\label{Sec:A}

We start with a two-level system coupled to two heat baths ($i=h,c$) comprising collections of harmonic oscillators ($\hbar= 1$, $k_B=1$),
\bea
H_{SSB} &=& \frac{\epsilon}{2}\sigma_z + \frac{\Delta}{2}\sigma_x + \sigma_z\sum_{k,i\in{\{h,c\}}}f_{k,i}(c_{k,i}^{\dagger}+c_{k,i}) 
\nonumber\\
&+& \sum_{k,i\in{\{h,c\}}}\nu_{k,i}c_{k,i}^{\dagger}c_{k,i}.
\label{eq:HSB}
\eea
Here, $\epsilon$ is the energy splitting of the spin, $\Delta$ is the tunneling frequency, 
$\sigma_{x,y,z}$ are the Pauli matrices, $c_{k,i}^{\dagger}$ ($c_{k,i}$) are the bosonic creation 
(annihilation) operators for the hot and cold baths ($i=h,c$) for a mode of frequency $\nu_{k,i}$. 
The displacements of the harmonic oscillators are coupled to the spin polarization with strength $f_{k,i}$,
and these interactions are captured by the spectral density functions, 
$J_{SSB,i}(\omega) = \sum_{k}f_{k,i}^2\delta(\omega-\nu_{k,i})$.
In what follows, we refer to the model Hamiltonian Eq. (\ref{eq:HSB}) as the ``standard spin-boson model" (SSB).

Technically, it is convenient to first specify the spectral function of the residual baths, after the RC 
mapping (as we do in Sec. \ref{Sec:B}), then work our way back and derive the spectral 
function of the SSB model.
For completeness, we already present here the spectral function of the SSB model:
It can be shown that an Ohmic function in the RC representation corresponds to a Brownian-oscillator
model for the SSB model, 
\begin{equation}
J_{SSB,i}(\omega) = \frac{4 \gamma_i\omega \Omega_i^2 \lambda_i^2}{(\Omega_i^2-\omega^2)^2 + (2\pi \gamma_i \Omega_i\omega)^2}.
\label{eq:JSB}
\end{equation}
We elaborate on this mapping in Appendix \ref{app:1}.
The physical picture emerging from this spectral function is that, in each bath,
there is a collection of modes with frequencies centered around
$\Omega_i$ that are strongly coupled to the system.
The energy parameter $\lambda_i$ quantifies the coupling strength of the spin to the $i$th bath.
The combination $\gamma_i\Omega_i$, with $\gamma_i$ a dimensionless parameter,
provides a measure for the width of the spectral function.
Since bath modes around $\Omega_i$ dominate the Brownian spectral function,
by extracting an oscillator around this frequency 
from the $i$th bath, and including it into the system, 
we can capture strong system-bath coupling effects.

In simulations below, we work with the following range of parameters:
$\Delta=1$,  $\epsilon=0$ (for simplicity), 
$\Omega > \Delta$, small width for the spectral function $\gamma\ll 1$, 
and variable system-bath coupling strength $\lambda/\Delta$.

\subsection{Reaction Coordinate mapping}
\label{Sec:B}

We follow a procedure analogous to that described in Ref. \citep{NazirPRA14}, 
but extended here to the case of a system strongly coupled to {\it two} heat baths. 
In brief, we apply a normal-mode transformation to the harmonic part of Eq. (\ref{eq:HSB})
and define a collective coordinate for each environment, termed reaction coordinate.
The two RCs ($i=h,c$) are each coupled directly to the spin, as well as to their residual harmonic bath. 
This mapping is sketched in Fig. \ref{fig:RC_TLS}.
The resulting reaction coordinate (RC) Hamiltonian is given by
\bea
H_{RC} &=& \frac{\epsilon}{2}\sigma_z + \frac{\Delta}{2}\sigma_x 
+\sum_i \Omega_i a_i^{\dagger}a_i
+ \sigma_z \sum_{i} \lambda_i (a_{i}^{\dagger} + a_i) 
\nonumber\\
&+&  \sum_{k,i}\omega_{k,i}b_{k,i}^{\dagger}b_{k,i} 
 + \sum_i (a_i^{\dagger}+a_i)\sum_k g_{k,i} (b_{k,i}^{\dagger}+b_{k,i})
\nonumber\\
&+& \sum_i(a_i^{\dagger} + a_i)^2\sum_k\frac{g^2_{k,i}}{\omega_{k,i}}.
\label{eq:HRC}
\eea
The two collective coordinates are defined such that,
\bea
\lambda_i(a_i^{\dagger} +a_i) = \sum_{k}f_{k,i} \left ( c_{k,i}^{\dagger}+c_{k,i}\right).
\eea 
Here, $\lambda_i$ is the newly-defined coupling of each RC to the spin system, 
$\Omega_i$ are the corresponding frequencies of the reaction coordinates. 
The residual baths are denoted by creation (annihilation) 
operators $b_{k,i}^{\dagger}$ ($b_{k,i}$), which now couple only to the $i$th RC. 
The spectral density functions of the residual heat baths are defined as
$J_{RC,i}(\omega)=\sum_{k} g_{k,i}^2\delta(\omega-\omega_{k,i})$. 

It is important to note that this mapping is exact. 
To complete the mapping, the spectral density functions of the SSB and the RC models 
need to be coordinated such that the spin system follows a corresponding dynamics.
For simplicity, we assume that the spectral functions of the residual baths are Ohmic,
\begin{equation}
J_{RC,i}(\omega) = \gamma_i\omega e^{-\omega/\Lambda_i}.
\label{eq:JRC}
\end{equation}
Here, $\gamma_i$ is a dimensionless coupling parameter of the RC to the residual bath.
This parameter is assumed to be small in our simulations allowing us to use the second-order QME
for the model after the RC mapping. $\Lambda_i$ is the cutoff frequency of the bath.

The reaction coordinate Hamiltonian, Eq.  (\ref{eq:HRC}), comprises the following components: 
The enlarged system (supersystem), now including the original spin, 
two additional harmonic oscillators as the RCs, 
and the coupling of the spin to these two oscillators. 
Other terms in Eq. (\ref{eq:HRC}), appearing in the second line
are the two residual heat baths ($i=h,c$) and their couplings to the two reaction coordinates.
The last element (third line) is a nontrivial quadratic term,
$\sum_i \left( a_i^{\dagger} +a_i\right)^2\sum_{k}\frac{g_{k,i}^2}{\omega_{k,i}}$. 
In Ref. \cite{NazirPRA14} it was treated as part of the interaction Hamiltonian and 
shown to be partially eliminated from the Markovian quantum master equation.
In Ref. \cite{Correa19}, the exactly solvable model of a chain of harmonic oscillators coupled to 
harmonic baths was treated with the Langevin Equation, then compared to the approximate Lindblad method.
It was pointed out that standard QME methods (such as
in Lindblad form or Redfield) have to be executed 
without the quadratic term. 

Essentially, an exact quantum equation of motion
for the Hamiltonian (\ref{eq:HRC}) is reduced to a Markovian QME in which the bare parameters of the RC
and its coupling to the system and bath 
($\Omega$, $\lambda$, $g_{k,i}$) are to be used, rather than renormalized parameters defined once
the quadratic term is absorbed \cite{YanRev}.
%
%
In Appendix \ref{app:2} we perform  a rotation
transformation and absorb the quadratic term, by renormalizing the parameters of the 
reaction coordinate Hamiltonian.
However, as pointed out in Refs. \cite{YanRev,NazirPRA14,Correa19},
Markovian QME simulations in this rotated basis do not correspond
to the original Hamiltonian. 
\subsection{Truncation of the RC Hamiltonian} 
\label{Sec:D}

Simulations of the RC model, Eq. (\ref{eq:HRC}) are performed by truncating it
to include only $M$ equi-distant levels. 
We count these levels using an index $l$,
\bea
\Omega a^{\dagger} a &\xrightarrow[]{}&\sum_{l=0}^{M-1} 
\Omega\left(l + \frac{1}{2}\right)
|l\rangle \langle l|,\,\, 
\nonumber\\
\left(a^{\dagger} + a\right)
& \xrightarrow[]{} &\sum_{l=1}^{M-1}\sqrt{l}\left( |l\rangle \langle l-1| + |l-1\rangle \langle l| \right). 
 \eea
For simplicity, we include the same number of states for each truncated RC.
After the truncation, 
the Hamiltonian  describes an enlarged system (ES) with a spin coupled to two $M$-level manifolds, 
with each $M$ manifold coupled to a residual heat bath,
%
\bea
H_{RC}^{M}= H_{ES}^{M} + H_B + H_{ES,B}^M.
\label{eq:HRCnM}
\eea
Explicitly, the enlarged system Hamiltonian is 
\bea
H_{ES}^M
&=&\frac{\epsilon}{2}\sigma_z + \frac{\Delta}{2}\sigma_x 
 +\sum_{i=h,c}\sum_{l_i=0}^{M-1} \Omega_i \left(l_i+\frac{1}{2}\right) |l_i\rangle \langle l_i|
\nonumber\\
&+&
\sigma_z \sum_{i=h,c} \lambda_i\sum_{l_i=1}^{M-1}\sqrt{l_i}|l_i\rangle \langle l_i-1| +h.c.
\label{eq:HESm}
\eea
%
As before, the residual baths are given as a collection of harmonic oscillators, 
$H_{B}= \sum_{k,i}\omega_{k,i}b_{k,i}^{\dagger}b_{k,i}$.
The extended system interacts with the two heat baths ($h,c$), 
\bea
H_{ES,B}^M= \sum_{i=h,c} V_{ES,i}^M \otimes B_i,
\eea
where we identify the system operator $V_{ES,i}^M$ and the coupled bath operator $B_i$ as
\bea
V_{ES,i}^M  &=& 
\sum_{l_i=1}^{M-1} \sqrt{l_i}\left(|l_i\rangle\langle l_i-1| +  |l_i-1\rangle\langle l_i|\right),
\nonumber\\
B_i&=&\sum_k g_{k,i} \left(b_{k,i}^{\dagger}+b_{k,i} \right).
\label{eq:HESBM}
\eea
The Hamiltonian $H_{ES}^M$ and system's operators $V_{ES,i}^M$ are 
matrices of dimension $2M^2 \times 2M^2$.


\subsection{Diagonalization and the effective spin-boson model}
\label{Sec:E}

\begin{figure}[htbp]
\centering
\includegraphics[width=8.55cm]{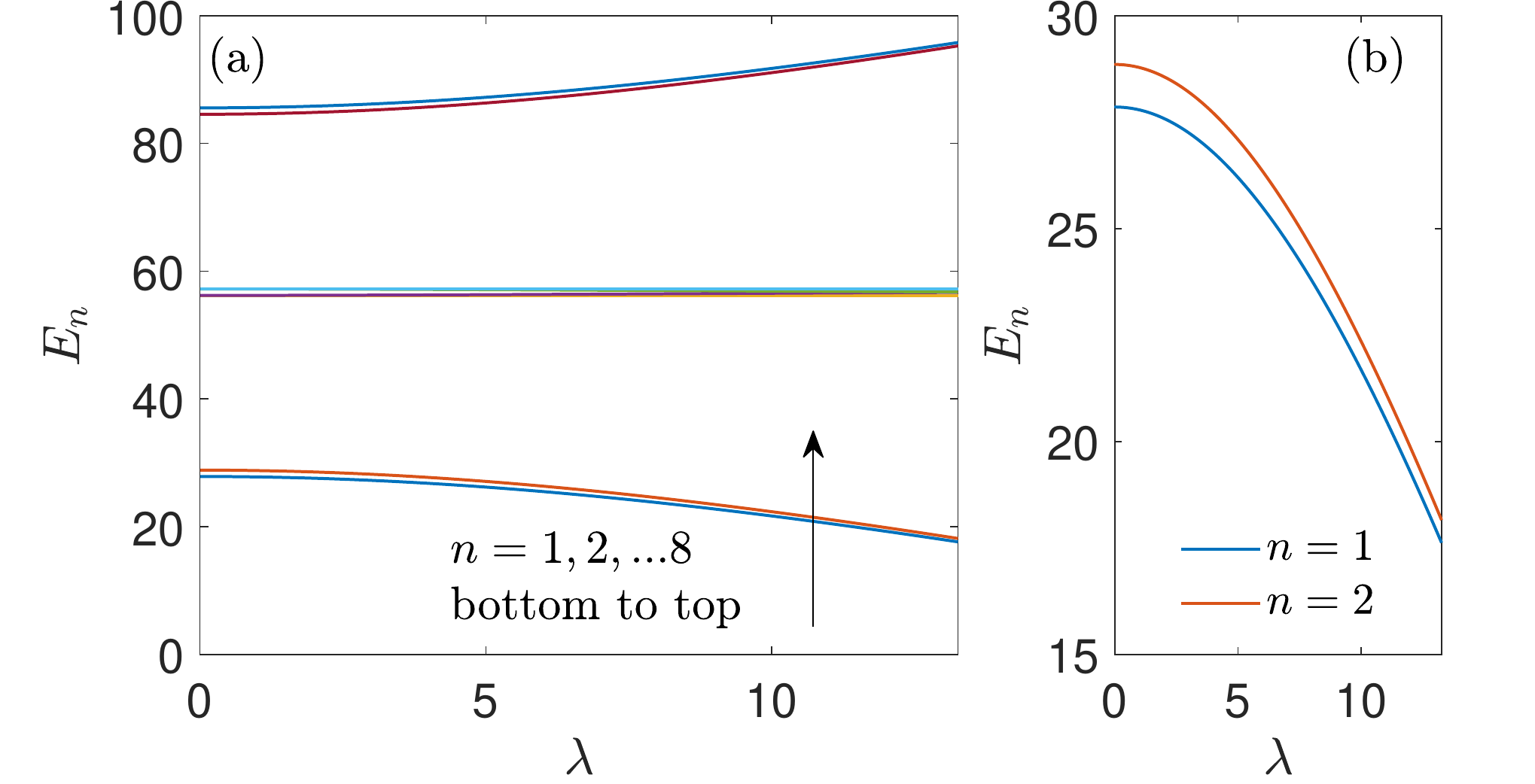} 
\caption{(a) Eigenenergies of $H_{ES}^{M=2}$
with $\Delta=1$, $\epsilon=0$, $\Omega=28 \Delta$ \cite{commentP}. 
(b) Focus on the lowest two eigenvalues, which form an effective spin Hamiltonian.
}
\label{Fig:spectrum}
\end{figure}

As a quick review, we started from the SSB model in Sec. \ref{Sec:A}. 
We extracted RCs, one from each bath in Sec. \ref{Sec:B} and 
we truncated the RCs in Sec. \ref{Sec:D}.
The last two steps of the procedure are described in this Section. They 
involve: (i) Diagonalizing the $M$-level RC extended-system $H_{ES}^M$ of Eq. (\ref{eq:HESm}).
(ii) Truncating the resulting spectrum, leaving only two states---thus strikingly, finishing where
we started with a two state system coupled to harmonic oscillator heat baths. 
Albeit, as we show next,
the spin-boson model at the end of the procedure relates to the SSB model through 
renormalized parameters.
In more details:

We diagonalize  $H_{ES}^M$ of Eq. (\ref{eq:HESm}) 
with a unitary transformation $U$. 
In general, this procedure is done numerically.
This transformation is further applied onto the interaction Hamiltonians in Eq. (\ref{eq:HESBM}),
\bea
H_{ES}^{D}= U^{\dagger} H_{ES}^MU, \,\,\,\,\
V_{ES,i}^D = U^{\dagger} V_{ES,i}^MU.
\eea
As an example, in Fig. \ref{Fig:spectrum} we present the evolution of the eigenenergies $E_n$
of $H_{ES}^M$ with the coupling energy, $\lambda$.
For simplicity, we use RCs with $M=2$, that is, the model includes a central spin and two
two-level RCs. Overall, there are eight (many-body) occupation-basis states.
For simplicity, the model is made to be fully symmetric; $\lambda=\lambda_i$ and $\Omega=\Omega_i$.

Since we work with $\Omega> \lambda$ and $\Omega \gg \Delta$,
the global eigenstates of $H_{ES}^M$  can be associated
with the local picture of a central spin and two (left and right) RC oscillators.
The lowest two eigenstates in Fig. \ref{Fig:spectrum}(b), are separated by 
$\approx \Delta$. They correspond to
the two RCs residing in their respective ground states 
(energy $2\times \frac{1}{2} \Omega$), while the central 
spin occupies either its ground or excited states.
Higher eigenenergies in Fig. \ref{Fig:spectrum}(a)
correspond to excited RCs, and are separated by multiples of $ \Omega$.
Specifically, there are four levels clustered around  $2 \Omega$ 
corresponding to one RC in its ground state and the other in its first excited state, 
with the spin occupying its ground or excited state.
The top two levels in Fig. \ref{Fig:spectrum} are associated with both RCs being excited,
and the spin occupying its ground or excited state.
One can similarly explain level structure for $M>2$.

\begin{figure}[htbp]
\includegraphics[width=8.55cm]{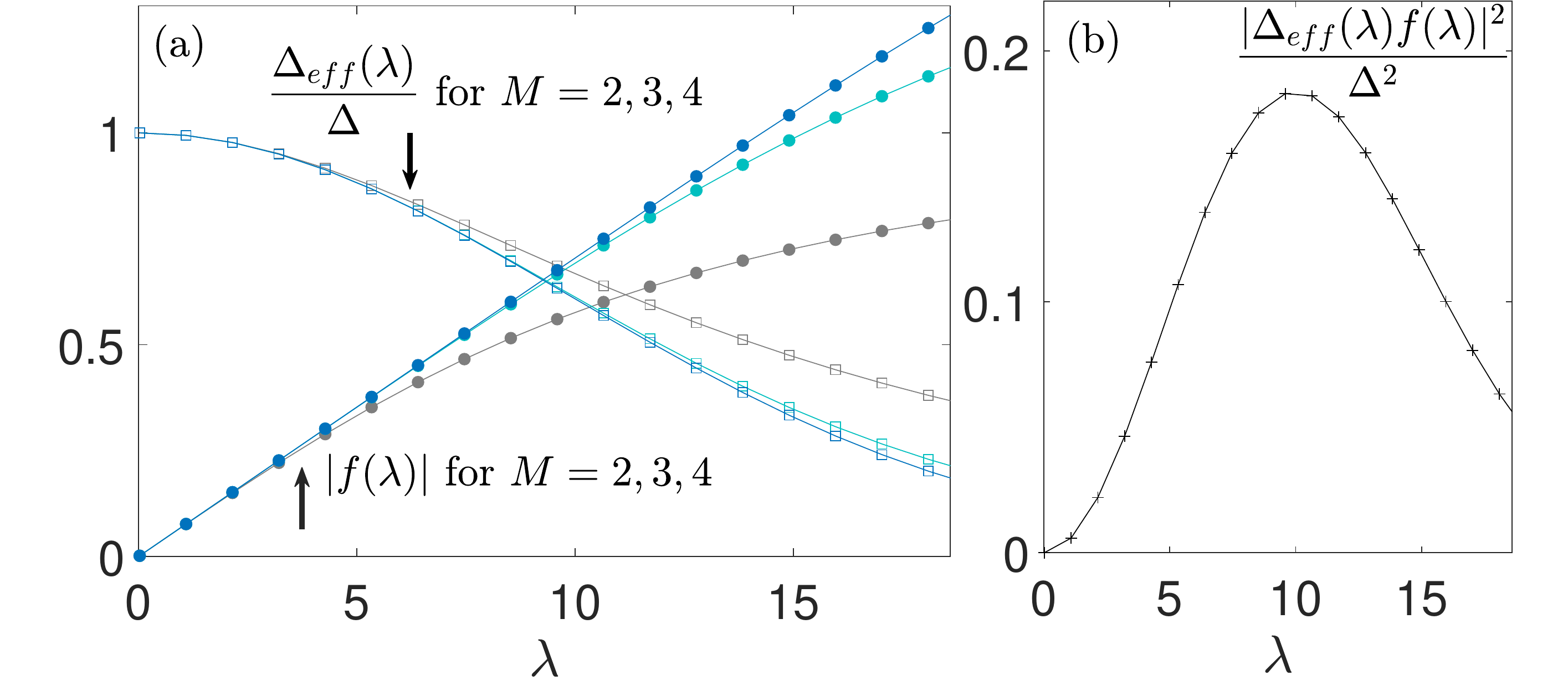} 
\caption{Behavior of the renormalized parameters in 
the effective spin-boson model, Eq. (\ref{eq:HSBeff}).
(a) The spin splitting $\Delta_{eff}(\lambda)$ and the coupling function $f(\lambda)$ are obtained numerically from the diagonalization transformation of $H_{ES}^M$.
(b) We present the combination $[\Delta_{eff}(\lambda)f(\lambda)]^2$, which controls the heat current
for ohmic spectral functions at weak coupling. 
$\epsilon$ = $0$, $\Omega$ = $28\Delta$.  
}
\label{effTLS}
\end{figure}

We now discuss our choice of parameters.
In previous studies e.g. in Ref. \cite{NazirPRA14},
the parameters of the RC model Eq. (\ref{eq:HRC})
were such that $\Omega\sim \Delta$, and $\Omega\sim T$.
Here, with the objective to uncover the role of strong couplings, we
use a high frequency RC mode,
$\Omega \gg \Delta$ and $\Omega \gg T$
such that we can safely truncate the spectrum of each RC and include  few levels when performing heat transport calculations.
This choice of parameters
allows us to focus on the strong-coupling limit while bypassing issues of
convergence involved in the truncation of the harmonic oscillator manifold.
As a result of this truncation, we are able to construct an effective model, which exposes
the essence of strong interactions, and is highly accurate as we 
show in section \ref{sec-simulA}.
In Appendix \ref{app:4} we include additional simulations of
the heat current at strong coupling using smaller values of $\Omega$,
down to $\Omega=5\Delta$.

Inspecting Fig. \ref{Fig:spectrum}, we make a useful observation: 
Since in simulations we work with $T\ll \Omega$,
excited states beyond the first two levels are thermally inaccessible as
there is a dramatic energy separation 
between the first two levels (gapped by $~\Delta$), and the rest of the eigenenergy manifold.
Based on this observation, we truncate the spectrum of $H_{ES}^D$ 
and write down an effective spin-boson Hamiltonian, 
\bea
H_{SB}^{eff} &=& \frac{1}{2}\Delta_{eff}(\vec \lambda)\sigma_z + \sigma_x \sum_if_i(\vec \lambda)\sum_k g_{k,i} (b_{k,i}^{\dagger} + b_{k,i})
\nonumber\\
&+& \sum_{k,i}
\omega_{k,i} b_{k,i}^{\dagger}b_{k,i},
\label{eq:HSBeff}
\eea
%
with
\bea
\Delta_{eff}(\vec \lambda)=E_{2}(\vec\lambda)-E_{1}(\vec\lambda), \,\,\,\,\,
f_i(\vec\lambda)  = \langle 1(\vec\lambda)| V_{ES,i}^D|2(\vec\lambda) \rangle.
\nonumber\\
\eea
We denote the pair $\lambda_i$ with a vector notation,
${\vec \lambda}=(\lambda_h, \lambda_c)$.
$E_n(\vec\lambda)$ are the eigenvalues of $H_{ES}^D$ with eigenstates 
$|n(\vec\lambda)\rangle$; $n=1,2$, as depicted e.g. in Fig. \ref{Fig:spectrum}.
We refer to the Hamiltonian (\ref{eq:HSBeff}) as the ``effective spin-boson (EFF-SB) model".

We highlight the dependence of the spin splitting on the interaction energies $\vec\lambda$.
The dimensionless function $f_i(\vec\lambda)$ arises from the transformation of the enlarged system Hamiltonian into the diagonal basis and it describes the dressing of the system-bath coupling energies. It depends
on both coupling energies, $\lambda_h$ and $\lambda_c$ and
this fact immediately suggests that cooperative, two-bath interactions should play a role in the dynamics and steady state transport behavior, 
as indeed is the case based on the polaronic approach \cite{PRL05, QME06,NIBA11,NIBA14}. 

To understand the impact of strong system-bath couplings, we depict in 
Fig. \ref{effTLS} the effective spin splitting 
and the coupling function $f_i(\vec\lambda)$. 
Since henceforth we use equal couplings (until Sec. \ref{sec-simulB}),  we relieve the vector notation, and identify by 
$\lambda$ the (equal) couplings of the spin to both reservoirs. 
On the one hand, strong coupling effects should facilitate transport by enhancing
the coupling of the spin to the baths. Indeed, $f(\lambda)$ is an increasing function of $\lambda$.
On the other hand, the detrimental impact of strong coupling  clearly emerges through the suppression of $\Delta_{eff}(\lambda)$, 
the tunneling splitting between the levels of the effective spin. 
Roughly, as we reduce the spin splitting, less energy is transmitted per quanta transferred.
In Appendix \ref{app:3} we present
analytical derivations on a simpler model to support the observed suppression of $\Delta_{eff}$
with $\lambda$.
When focusing on quantum dynamics, the suppression of $\Delta_{eff}(\lambda)$ would 
eventually lead to the localization of the spin \cite{Weiss}. 
In calculations of steady state heat transport as we perform here, 
the substantial suppression of $\Delta_{eff}(\lambda)$ shows up in the 
rapid decline of the heat current at strong coupling. 

We fit the functions in Figure \ref{effTLS}(a) and find that, approximately, 
$\Delta_{eff}(\lambda)\propto \exp\left(-\frac{\lambda^2}{\lambda_m^2}\right)$ 
and $f(\lambda)\propto \lambda$, with $\lambda_m$ a parameter with the dimension of energy.
Importantly, since $f(\lambda)$ is of order 1, 
as long as $\gamma\ll1$ [see Eq. (\ref{eq:JRC})],
we can still safely employ a weak coupling scheme to treat the dynamics and transport characteristics of
the effective-SB Hamiltonian Eq. (\ref{eq:HSBeff}).

We present the combination $[\Delta_{eff}(\lambda)f(\lambda)]^2$ 
in Fig. \ref{effTLS}(b) and reveal a non-monotonic behavior: 
The product first grows approximately 
as $\lambda^2$ then it decays rapidly as $\exp(-\lambda^2/\lambda_m^2)$, 
with a maximum at $\lambda_m/\sqrt{2}$.
As we discuss in more details in Sec. \ref{sec-simulA},  
the steady state heat current in the SB model is proportional to this product
for Ohmic baths under the weak system-bath coupling approximation
\cite{PRL05,QME06,NIBA14,ARPC}. Thus, the trend observed in Fig. \ref{effTLS}(b) is representative
of the heat current in this regime.

While the condition of weak system-bath coupling does not hold for the SSB model since 
$\lambda/\Delta$ can be large,
{\it it is valid for the effective SB model} since its coupling to the bath is dictated by the width
parameter ${\gamma}_i$, which is assumed small.
This is the main physical and computational outcome of the mapping approach:
The effective spin-boson model 
captures strong coupling effects, as inherited from the SSB model,
but now embedded in renormalized parameters while allowing a weak coupling treatment to the baths.

The effective SB model described in this subsection is highly beneficial
as it hands over analytic understanding on transport characteristics.
It is important to appreciate that the EFF-SB model should be identified
only after constructing a RC that converges with respect to $M$, the number of levels representing the RC.
This aspect is clearly manifested in Fig. \ref{effTLS}.
While $\Delta_{eff}(\lambda)$ and $f(\lambda)$ readily converge with $M$ at weak-intermediate 
couplings, at strong coupling increasingly more levels should be included to converge the parameters of the effective SB model.


\section{Method: steady state energy transport with the RC-QME}
\label{sec-Redfield}

We succeeded in transforming the SSB model with possibly strong coupling of the spin to harmonic oscillator baths into an effective SB model. 
The parameter $\gamma\Omega$ in the SSB model, which controls the width of the spectral function,
dictates the system-bath coupling energy in the RC picture.
Thus, if we assume a narrow spectral function for the SSB model, in the RC model the extended system becomes weakly coupled to its surroundings.
As a result, the dynamics and steady state characteristics of the effective SB can 
be studied using a QME perturbative in the system-bath coupling. 

Here, we employ the Markovian Redfield master equation to study the steady state behavior under the 
Hamiltonian (\ref{eq:HRCnM})-(\ref{eq:HESBM}).
This approach relies on three assumptions: 
(i) The enlarged system is weakly coupled to its residual environments
such that a second-order perturbative approach yields proper results. 
(ii) The residual environments act as Markovian baths. 
(iii) The initial state of the total system 
factorizes into a system times bath form.
The residual thermal baths are individually prepared in their canonical states characterized by inverse temperatures $\beta_i$.

Working in the Schr\"odinger picture and in the energy basis of the subsystem, the Redfield equation for the reduced density matrix of the enlarged system (dimension $2M^2$) is given by \cite{Nitzan,Breuer},
\bea
\dot{\rho}_{ES,mn}(t) &=& -i\omega_{mn} \rho_{ES,mn}(t) 
\nonumber\\ &+&
\sum_i\sum_{j,k}[
R_{mj,jk}^{i}(\omega_{kj}) \rho_{ES,kn}(t) + R_{nk,kj}^{i*}(\omega_{jk}) \rho_{ES,mj}(t)
\nonumber\\ &-& 
R_{kn,mj}^{i}(\omega_{jm}) \rho_{ES,jk}(t) - R_{jn,mk}^{i*}(\omega_{kn}) \rho_{ES,jk}(t)].
\label{eq:RMEQ}
\eea
Here, $\omega_{mn}=E_m-E_n$ are the bohr frequencies of the enlarged system in the energy basis.
The $R$ super-operator terms in Eq. (\ref{eq:RMEQ}) are given by half Fourier transform of bath autocorrelation functions,
\bea
R_{mn,jk}^{i}(\omega) &=& (V_{ES,i}^{D})_{m,n} (V_{ES,i}^{D})_{j,k} \int_0^{\infty} d\tau e^{i\omega\tau} \langle B_i(\tau)B_i\rangle
\nonumber\\ &=&
(V_{ES,i}^{D})_{m,n} (V_{ES,i}^{D})_{j,k} [\Gamma_{RC,i}(\omega) + i\Delta_{RC,i}(\omega)].
\label{eq:dissipator}
\eea
The autocorrelation functions of bath operators are evaluated with respect to the 
thermal state of their respective baths.
In what follows, we neglect the imaginary component of the dissipators, $\Delta_{RC}$, as it can be shown that they cancel out the quadratic term in the RC Hamiltonian, Eq.(\ref{eq:HRC}) \cite{NazirPRA14}.

For harmonic environments and a bilinear coupling between the enlarged system and the
residual environments, the real part of the $R$ term reduces to
\bea
\Gamma_{RC,i}(\omega) = 
\begin{cases}
\pi J_{RC,i}(\omega) n_i(\omega) & \omega > 0   \\
\pi J_{RC,i}(|\omega|)[(n_i(|\omega|) + 1] & \omega < 0.
\end{cases}
\eea
Here, $n_i(\omega)$ is the Bose-Einstein distribution function of the $i$th bath; 
$J_{RC,i}(\omega)$ is the RC spectral function given by Eq. (\ref{eq:JRC}). 

To calculate the heat current flowing through the system, 
we write down the Redfield equation as
\bea
\dot \rho_{ES}(t)=-i[H_{ES}^{D},\rho_{ES}] + D_h(\rho_{ES}) + D_c(\rho_{ES}),
\eea
with the dissipators $D_{h,c}(\rho_{ES})$ organized based on Eq. (\ref{eq:RMEQ}). The dissipators are additive given the weak coupling of the RCs to their residual baths
(but each dissipator $D_i$ depends non-additively on the original coupling parameters, $\lambda_i$). 

We solve the equation of motion in steady state, $\dot{\rho}_{ES}(t) = 0$ and obtain 
the density matrix of the enlarged system, $\rho_{ES}^{ss}$.
The steady state heat current at the $i$th bath, calculated from the heat exchange between the  enlarged system and this reservoir is given by
\bea
J_i={\rm Tr}\left[ D_i(\rho_{ES}^{ss}) H_{ES}^{D}\right].
\eea 
The heat current is defined positive when flowing from the $i$th bath towards the system.
This procedure, the application of the Redfield-QME onto the RC model
is referred to as the reaction-coordinate quantum master equation method, RC-QME. 

Below, we perform heat transport simulations  of the nonequilibrium spin-boson model 
using four methods:

{\bf (i) RC-QME framework}: We follow the approach described in this Section,
while checking convergence by increasing $M$, 
the number of levels used to describe the RCs.

{\bf (ii) Weak coupling, BMR-QME calculations}:
We calculate the heat current with the original SSB Hamiltonian Eq. (\ref{eq:HSB}) 
without performing the RC mapping
under the (unjustified) weak coupling assumption, so as to highlight signatures of strong-couplings.
In this case, we again employ the Redfield Equation Eq. (\ref{eq:RMEQ}) in the diagonal representation 
but  use  
$J_{RC}(\omega) \xrightarrow{} J_{SSB}(\omega)$. 
%
Obviously, this calculation (based on an analytic expression \cite{PRL05,QME06,NIBA14,ARPC})
is appropriate  only when the coupling of the spin to the baths is weak. 

{\bf (iii) PT-NEGF}: We use the polaron-transformed 
NEGF method developed in Ref. \cite{PT-NEGF17}, denoted by ``PT-NEGF".
This method, tested on the nonequilibrium spin-boson model,
is capable of treating nonperturbatively the system-bath coupling term.
It also handles  the tunneling energy in a perturbative manner.
The method combines the polaron transformation with the nonequilibrium Green's function method, where a
Majorana-fermion representation is utilized to evaluate the Dyson series.

{\bf (iv)  EFF-SB model}:  
We obtain the parameters of the effective two-state 
model based on the diagonalization of $H_{ES}^M$ with large enough $M$.
However, since we end up with two states weakly coupled to the surroundings,
we can utilize the closed-form expression for the heat current as derived in Refs.
\cite{PRL05,QME06}, assuming a Markovian, Ohmic bath.
We write down this expression emphasizing the nontrivial dependence of parameters on the renormalized spin splitting,
\begin{widetext}
\bea
J_q^{eff}=
\Delta_{eff} \frac{   J_{RC,h}^{eff}(\Delta_{eff})    
J^{eff}_{RC,c}(\Delta_{eff})
\left[ n_h(\Delta_{eff})- n_c(\Delta_{eff})\right]}
{J^{eff}_{RC,h}(\Delta_{eff})  \left[  2n_h(\Delta_{eff})+1 \right] + J^{eff}_{RC,c}(\Delta_{eff})  \left[  2n_c(\Delta_{eff})+1 \right]}.
\label{eq:jqeff}
\eea
\end{widetext}
%
The Ohmic spectral function, which is evaluated at the renormalized splitting, is dressed by $f(\lambda)$.
For brevity, we suppress the dependence of $\Delta_{eff}$ on $\lambda$,
\bea J^{eff}_{RC}(\Delta_{eff}) = \gamma \Delta_{eff} |f(\lambda)|^2 e^{-\Delta_{eff}/\Lambda}.
\eea
For a symmetric system in the limit of high $\Lambda$ the heat current reduces to 
\bea
J_q^{eff}  = \gamma |f(\lambda)|^2 [\Delta_{eff}(\lambda)]^2 \frac{n_h(\Delta_{eff}) - n_c(\Delta_{eff})}
{2n_h(\Delta_{eff}) + 2n_c(\Delta_{eff})+2}.
\nonumber\\
\label{eq:jqeffs}
\eea
In Fig. \ref{effTLS}(b) we showed that the temperature-independent prefactor of this function displayed a turnover behavior with $\lambda$,


\begin{figure*}[htb]
\vspace{-4mm}
\includegraphics[width=15cm]{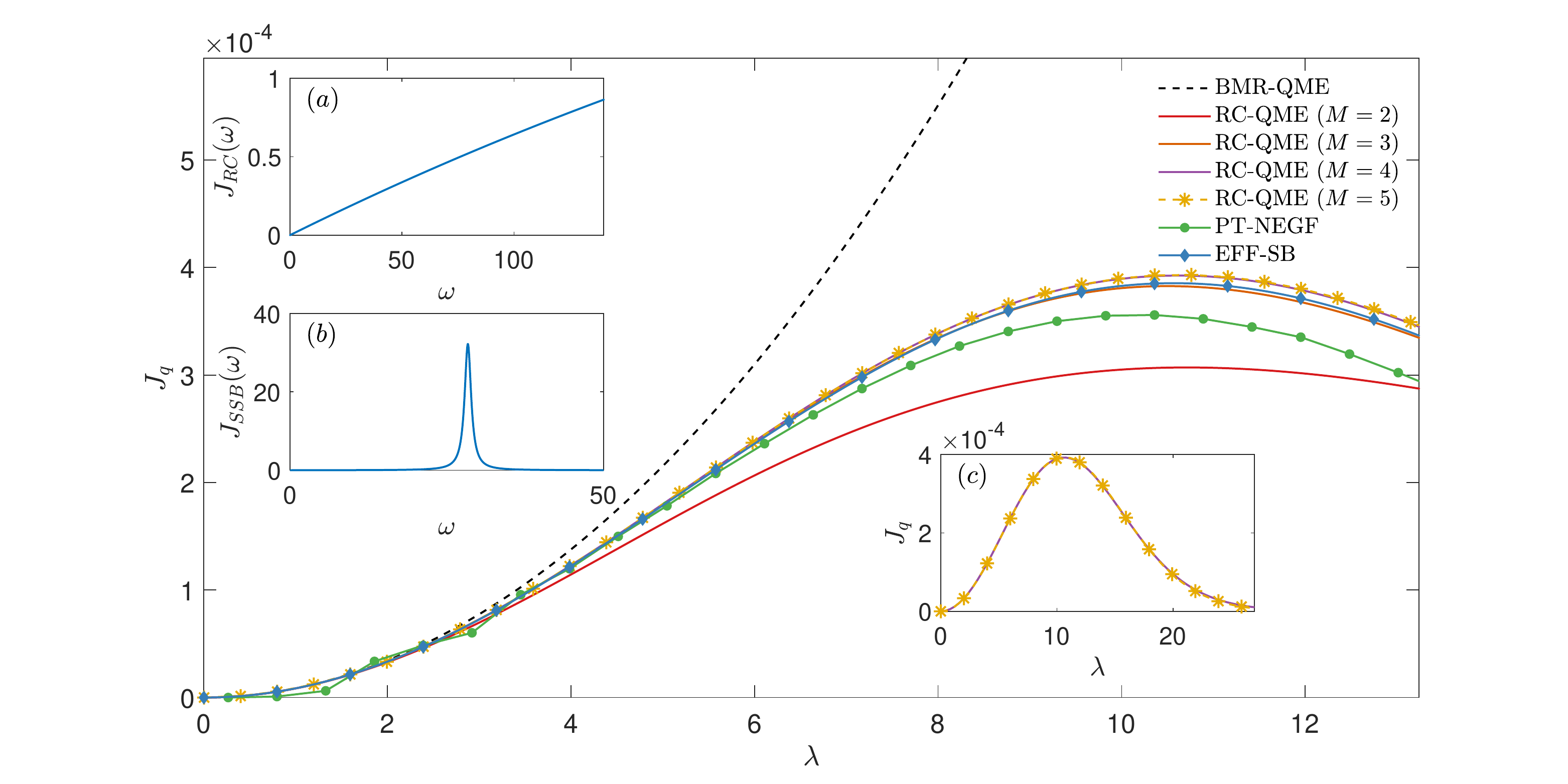} 
\caption{Steady state heat current as a function of coupling strength  $\lambda$.
We compare RC-QME simulations ($M=2-5$) to BMR-QME (dashed) and PT-NEGF ($\circ$) results.
RC-QME simulations with $M=4-5$ overlap.
We further display the current predicted from the
effective SB model (EFF-SB), Eq. (\ref {eq:jqeff}).
(a)-(b) Spectral density functions of the reaction
coordinate model and the standard spin-boson model, respectively.
Parameters are 
$\epsilon$ = $0$, $\Omega$ = $28\Delta$,  $\gamma$ = $0.0071/\pi$ and
$\Lambda=1000\pi\Delta$.  
The temperatures are $T_h$ = $\Delta$, $T_c$ = $0.5\Delta$.
In panel (b), we employ $\lambda$ = $8 \Delta$.
(c) Demonstration of the full suppression of the heat current at strong coupling based on RC-QME simulations for $M=4$ and $M=5$ (overlapping).
}
\label{fig:Current_vs_Coupling}
\end{figure*}


\section{Numerical Simulations}
\label{sec-simulA}


\subsection{System-bath coupling}

We begin by studying the behavior of the heat current as a function of 
the system-bath coupling strength.
In the SSB model of Sec. \ref{Sec:A}, the parameter $\lambda_i$ 
quantifies the system-bath coupling, see Eq. (\ref{eq:JSB}). 
In the RC picture, this parameter describes 
the coupling of the spin to the added reaction coordinates,
while the enlarged-system-bath coupling is controlled by $\gamma$.

Figure \ref{fig:Current_vs_Coupling} depicts the  steady state heat current 
as a function of $\lambda$ using our RC-QME framework.
We further show the BMR-QME, EFF-SB, and the PT-NEGF results.
For simplicity, we assume a symmetric setup,
$\lambda=\lambda_i$, $\Omega=\Omega_i$, $\gamma=\gamma_i$, $\Lambda=\Lambda_i$.
%

We find that the heat current obtained from the BMR-QME grows quadratically with $\lambda$;
recall that the spectral function scales as $J_{SSB}(\omega)\propto \lambda^2$.
In contrast, the RC-QME depicts a turnover behavior, in a good agreement with the PT-NEGF method
(which takes into account strong coupling effects through the Polaron transformation and the Dyson-series summation).
Notably, the turnover behavior of the current with the coupling energy 
is excellently captured by the effective-SB model,
Eq. (\ref{eq:jqeff}) with the renormalized parameters, displayed in Fig. \ref{effTLS}. 
%
Parameters for the effective model were taken using $M=4$. 

\begin{figure}[htbp]
\centering
\includegraphics[width=8.55cm]{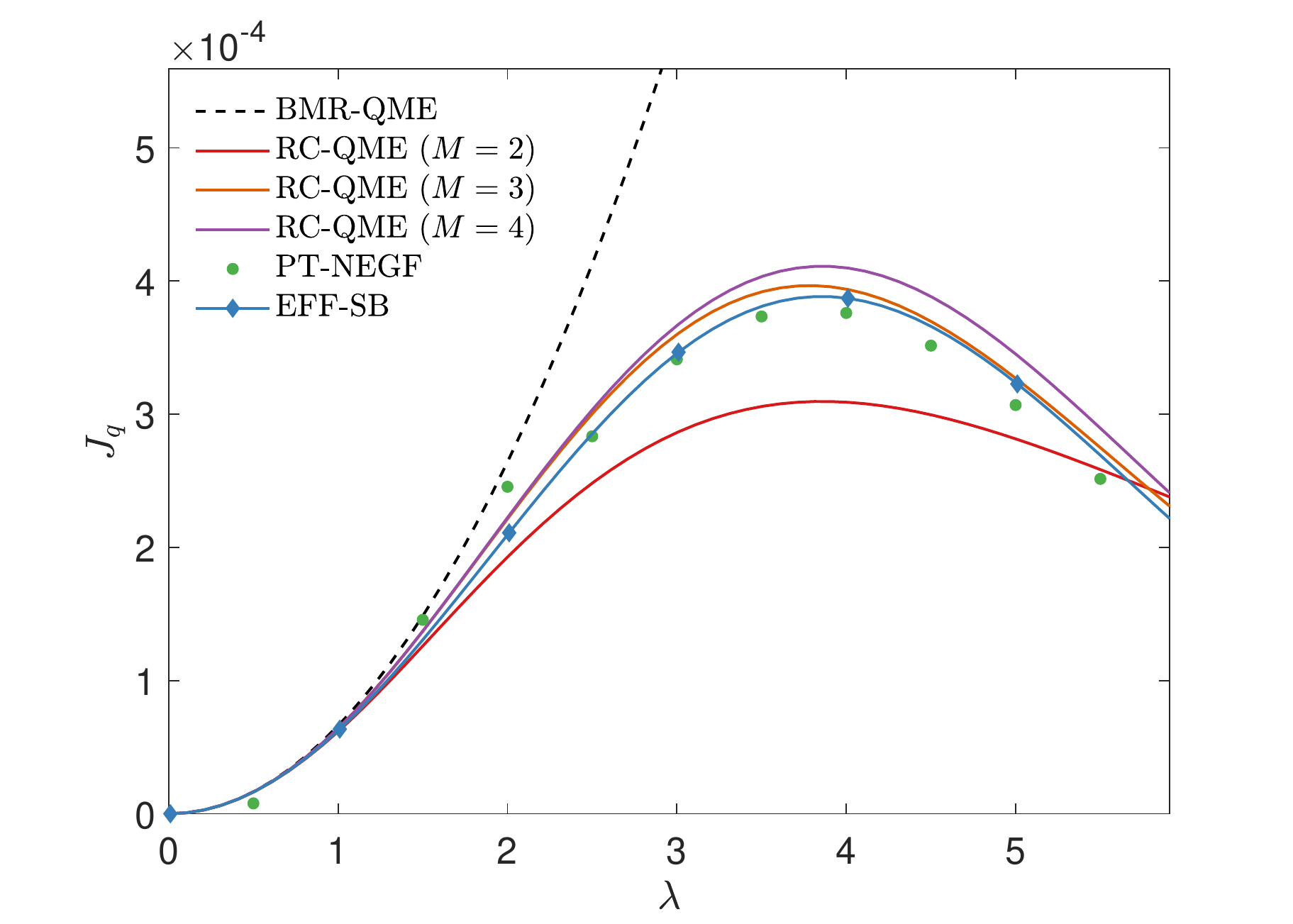} 
\caption{Steady state heat current as a function of coupling strength $\lambda$. 
Parameters are the same as in Fig. \ref{fig:Current_vs_Coupling}, except $\Omega = 10 \Delta$. 
}
\label{fig:Current_vs_lambda10}
\end{figure}

Figure \ref{fig:Current_vs_lambda10} shows the steady state heat current 
for a smaller value of $\Omega$.
 Recall, that this parameter controls the peak position and width
of the SSB spectral density, Eq. (\ref{eq:JSB}). 
We again observe a very good agreement between strong coupling methods.
Notably, we notice that the EFF-SB better agrees with the PT-NEGF technique than the RC-QME method
at $M=4$.
This trend is further exemplified in Appendix \ref{app:4}, and discussed in details in Sec. \ref{sec:excep}.


\begin{figure*}[htbp]
\centering
\includegraphics[width=17cm]{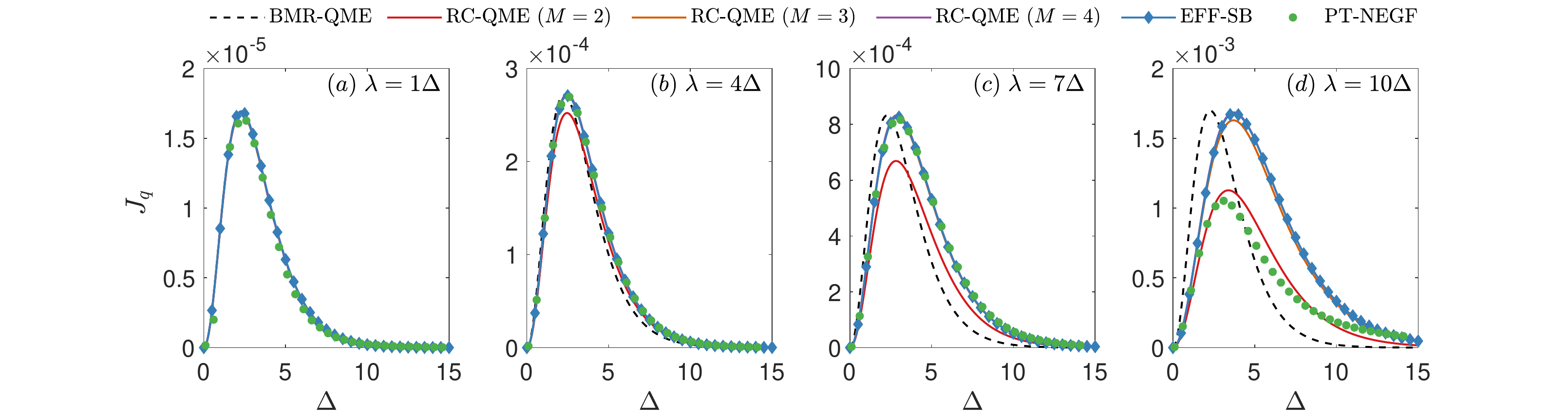} 
\caption{(a)-(d) Steady state heat current as a function of tunneling splitting 
from weak to strong couplings.
Parameters are the same as in Fig. \ref{fig:Current_vs_Coupling}. 
}
\label{fig:Current_vs_splitting}%
\end{figure*}

\subsection{Tunneling splitting}

We study the behavior of the heat current as a function of the tunneling 
splitting $\Delta$ in Fig. \ref{fig:Current_vs_splitting}, separately looking
at weak (a), intermediate (b)-(c) and strong (d) system-bath couplings.
%
%
The dependence of $J_q$ on the tunneling splitting was previously investigated analytically
with NIBA \cite{NIBA11} and the BMR-QME \cite{Nazim14}, and numerically, with exact 
methods \cite{Segal13,Nazim14,PI}.
The general trend observed is of the heat current first growing with $\Delta$ quadratically.
However, for larger spin splitting such that $\Delta>T$, $J_q$ begins to decay with $\Delta$.
This suppression of the current can be rationalized in the weak coupling picture 
when only resonant processes contribute:
When $\Delta$ becomes large, bath modes at this frequency are not thermally populated,
resulting in the suppression of the heat current.

Inspecting  Fig. \ref{fig:Current_vs_splitting}, we find that
the different methods agree at weak coupling.
However, as we move into the intermediate and strong coupling regimes 
the current predicted by the BMR-QME 
offsets from the RC-QME. 
This is expected: We learned from Fig. \ref{effTLS}
that strong coupling manifests itself in the renormalization (suppression) of the tunneling splitting.
As such, in the BMR-QME method the position of the peak occurs at $\Delta \sim T$ while it appears only
at a higher splitting for the RC-QME, once $\Delta_{eff} \sim T$. 
Interestingly, while the BMR-QME method incorrectly positions
the peak, the maximal magnitude of the current is the same as in the RC-QME method.

As for the PT-NEGF method, while it agrees with the RC-QME at weak-intermediate couplings, see
Fig. \ref{fig:Current_vs_splitting}(a)-(c), 
more substantial deviations show up at strong coupling (up to 50\%) and for large splitting, see
Fig. \ref{fig:Current_vs_splitting}(d).
However,  while the magnitude of the current differs between the RC-QME and the PT-NEGF methods,
the peak position agrees.
This manifests that the essential aspect of gap renormalization is captured in both methods, which is 
significant given the distinct ways they handle strong coupling effects.
%
Which method, RC-QME or PT-NEGF is more accurate at strong coupling and large spin-splitting? 
We think that the RC-QME provides a more accurate result in this difficult regime:
The PT-NEGF method does not handle exactly the spin splitting at large $\Delta$;
while it is more accurate than the Markovian NIBA method of Ref. \cite{NIBA11}, 
which can only capture the $J_q\propto \Delta ^2$ scaling,
the self-energy in the PT-NEGF includes $\Delta$ only perturbatively. 
In contrast, the RC-QME handles both $\Delta$ and $\lambda$ non-perturbatively.
Nevertheless, additional benchmarking is required to determine the exact
behavior of the current at strong coupling and at large tunneling splitting.

\begin{figure*}[htbp]
\centering
\includegraphics[width=17.90cm] {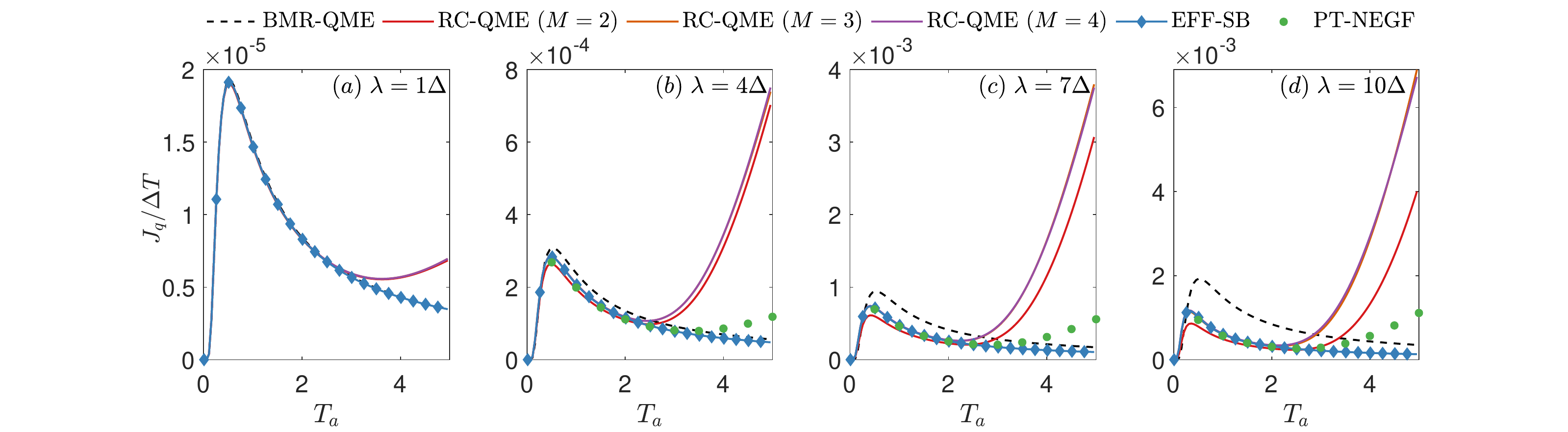} 
\caption{(a)-(d) Steady state heat current, divided by the temperature difference,
as a function of the average temperature of the baths.
Parameters are the same as in Fig. \ref{fig:Current_vs_Coupling}.
We set $\Delta T = 0.3T_a$, and $T_{h/c} = T_a \pm \Delta T/2$.}
\label{fig:Current_vs_temperature}%
\end{figure*}

\begin{figure}[htbp]
\centering
\includegraphics[width=8.55cm]{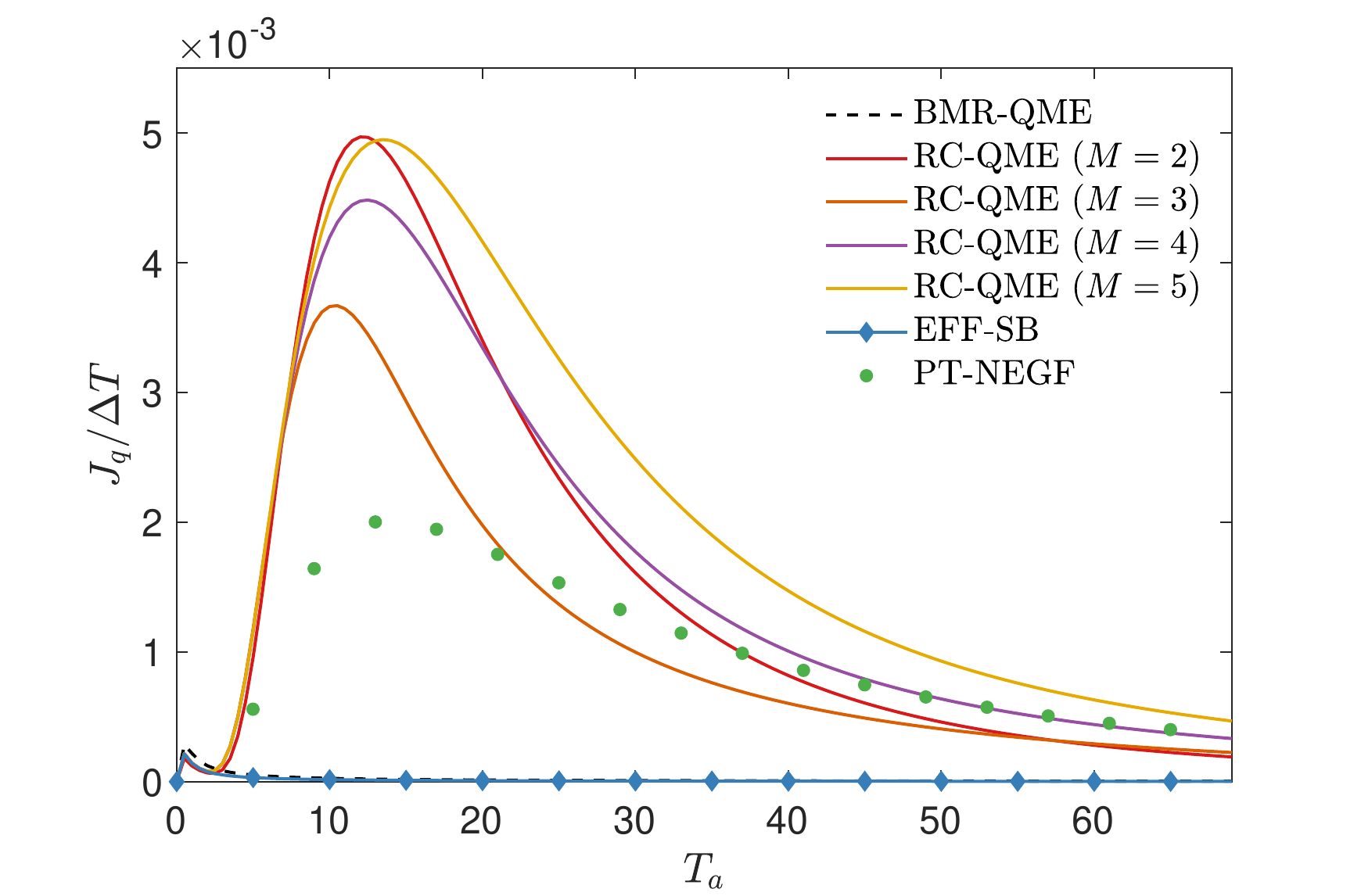} 
\caption{Steady state heat current divided by the temperature difference
as a function of the averaged temperature of the reservoirs.
This figure corresponds to Fig. \ref{fig:Current_vs_temperature} (c),
but continuing here into the high temperature regime.
Parameters are the same as in Fig. \ref{fig:Current_vs_Coupling}.}
\label{fig:Current_vs_temperature_large}
\end{figure}


\subsection{Temperature-difference dependence}

We study the behavior of the heat current 
as a function of the average temperature, $T_a \equiv (T_h + T_c)/2$
in Figs. \ref{fig:Current_vs_temperature}-\ref{fig:Current_vs_temperature_large}. 
Inspecting the low-intermediate temperature regime of $T_a<2\Delta$
in Fig. \ref{fig:Current_vs_temperature},
we observe that the different methods display a crossover behavior:
The current rises with $T_a$ as it approaches a resonance condition at $\Delta$ (or $\Delta_{eff}$ in the RC-QME).
However, at higher temperatures the current drops as $1/T_a$ due to the temperature dependence 
of the scattering rates as captured by Eq. (\ref{eq:jqeff}).
The BMR-QME overestimates the current at strong couplings since it does not capture the suppression
of $\Delta$, thus it allows for the transfer of more heat than in the RC-QME method.

As we raise the temperature, $T_a>2\Delta$, we observe 
in Fig. \ref{fig:Current_vs_temperature}, and in more details in Fig.
 \ref{fig:Current_vs_temperature_large},
a dramatic enhancement of the current, particularly at strong coupling. 
As a clue to deciphering the origin of this enhancement, 
we note that it is not captured by the EFF-SB model
thus it emerges from the contributions of higher-energy transitions 
in the manifold of the extended system, see Fig. \ref{Fig:spectrum}.
Naturally, we question whether the enhancement of the current at high temperature is physical,
or is it an artifact of the extraction of a single oscillator mode as a RC.
However, in Fig. \ref{fig:Current_vs_temperature_large} we show 
that the PT-NEGF method also displays this
enhancement, though there are differences in the magnitude of the current. 

Altogether, as a function of temperature, we gather from Figs. \ref{fig:Current_vs_temperature} 
and \ref{fig:Current_vs_temperature_large} that the heat current is peaked first around
$T_a \approx \Delta$   then at  $T_a\approx \Omega$ 
(there are additional peaks at higher temperature). 
In the RC picture, we can explain this structure as follows: 
The current is high when $T_a$ roughly matches transition frequency in the extended system.
The first peak corresponds to the excitation of the spin, with the RCs in their ground state.
The second peak corresponds to energy transfer via the excitation of the RCs.

Interestingly, using the PT-NEGF method 
we can explain these peaks with a different picture, 
thinking about resonant and off-resonant (tunneling) transport of phonons.
The first peak at $T_a\approx \Delta$ corresponds to resonant transmission of 
phonons from the bath through the spin.
Beyond that at higher temperatures, the transport process occurs increasingly 
via tunneling: high frequency phonons from the baths must utilize the low frequency spin impurity 
to transfer heat. This off-resonant process becomes more potent
as the temperature rises eventually matching 
the peak of the baths spectral density function (recall that we use a Brownian function in the SSB model), where the bath density of state is maximized.
If one continues and increases the temperature, additional peaks will show up at multiples of $\Omega$. They would represent multi-phonon transfer processes in the PT-NEGF method,
and the excitation of high energy levels in the extended-system RC model.


\subsection{Discussion: The exceptional performance of the effective Spin-boson model}
\label{sec:excep}

The effective SB model brings insight on the strong coupling regime, and it
allows fast computations.
More surprisingly, as we show e.g. in Figs. 
\ref{fig:Current_vs_Coupling}-\ref{fig:Current_vs_lambda10} and Appendix D, it consistently 
provided results more accurate than those achieved with
the full $2M^2$ model Hamiltonian, Eq. (\ref{eq:HRCnM}) as long as $T\ll \Omega$, $\Delta\ll \Omega$.
What is the explanation for the remarkable success of the EFF-SB model?
We can explain this superior behavior by noting that
weak-coupling methods typically overestimate the heat current compared to exact results \cite{Nazim14}.
In the RC-QME framework, as we include a large number of states for the RC ($M>2)$, 
many transitions contribute to the heat current, thereby 
increasing the error associated with the weak coupling scheme. 
Thus, while a large $M$ more accurately represents the RC oscillators, 
at the same time  errors associated with the calculation of the current using a weak-coupling method become more significant.
%

In contrast, the effective SB model offers an excellent, fortunate tradeoff of these two aspects:
On the one hand, it is constructed by diagonalizing a {\it large} $M$-RC Hamiltonian, 
Eq. (\ref{eq:HRCnM}); the renormalized parameters therefore capture the RC manifold.
On the other hand, the subsequent truncation of the $2M^2$ spectrum, to include only 
few (two) states limit the over-estimation error associated with weak coupling schemes.

\section{Thermal-diode effect}
\label{sec-simulB}
The nonequilibrium spin-boson model was originally introduced to 
describe a quantum thermal diode (rectifier) \cite{PRL05,QME06}.
In a boson-bath nanojunction, this effect relies on two conditions: anharmonicity, captured here
by the impurity spin, and spatial asymmetry, introduced here
by coupling the spin asymmetrically to the different baths.
Recently, the nonequilibrium spin-boson model was realized in a superconducting quantum circuit
demonstrating a diode effect with rectification ratio $R = |\frac{J_q(T_h,T_c)}{J_q(T_c,T_h)}| = 
1.1$ \cite{PekolaE}. 
An important question that has been touched on, without conclusion 
is whether strong coupling effects are beneficial for the diode effect, and to what extent.
The EFF-SB model offers computation at low cost, and we utilize it here to address this question.

The quantity of interest here is the thermal rectification ratio, 
$R$, which quantifies transport asymmetry when reversing the temperature bias. 
In this definition, $R = 1$ corresponds to lack of rectification, while $R > 1$, or $R < 1$ 
corresponds to a diode effect.
To materialize this effect, spatial symmetry needs to be introduced into the model. 
This is done here through the use of an asymmetry parameter $-1< \chi <1$,
with $\lambda_h = \lambda (1 - \chi)$, $\lambda_c = \lambda (1 + \chi)$. 
In this definition, $\chi = 0$ corresponds to the symmetric model 
where no rectification is expected.

Using Eq. \ref{eq:jqeff}, we obtain an analytic expression for the rectification ratio 
\bea
R &=& \frac{\frac{f_L(\vec\lambda)}{f_R(\vec\lambda)} (2n_c(\Delta_{eff})+1) + (2n_h(\Delta_{eff})+1) }{\frac{f_L(\vec\lambda)}{f_R(\vec\lambda)} (2n_h(\Delta_{eff})+1) + (2n_c(\Delta_{eff})+1)}.
\label{eq:Rectification}
\eea
This equation reveals that the ratio of the effective coupling parameters breaks spatial symmetry in the system, allowing a diode effect to emerge. 
Indeed, if $f_L(\vec\lambda) = f_R(\vec\lambda)$, then $R = 1$
and the system will not exhibit a diode behaviour.

\begin{figure}[htbp]
\vspace{2mm}
\hspace{-1mm}
\includegraphics[width=9cm]{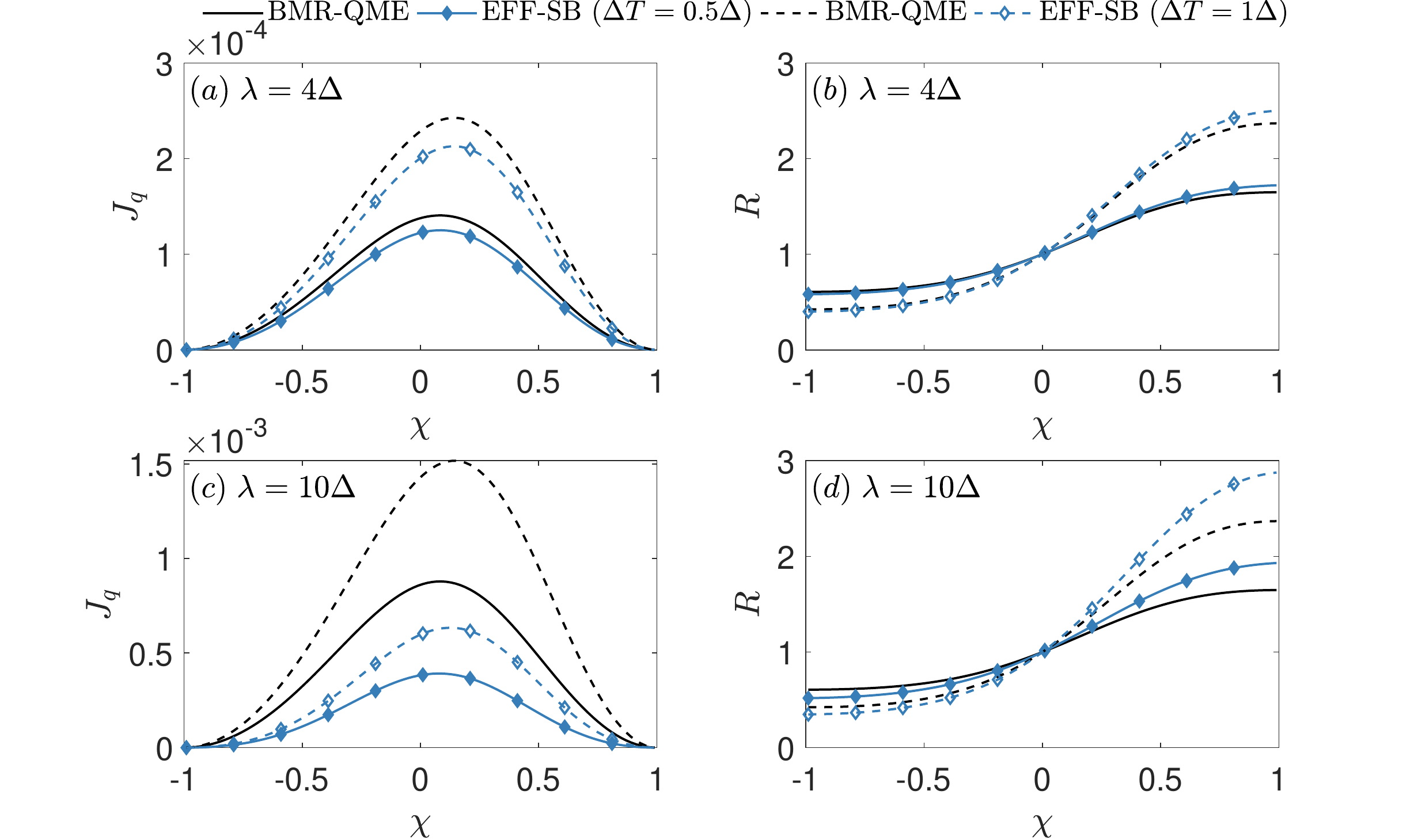} 
\caption{
Thermal diode effect at strong coupling with the EFF-SB model (filled and empty $\diamond$) 
and the weak coupling BMR-QME (full and dashed lines) at $\Delta T=0.5\Delta$ and $\Delta T=1\Delta$.
Panels (a) and (c) depict the steady state heat current as a function of the asymmetry parameter 
$-1<\chi<1$, $\lambda_h = \lambda (1 - \chi)$, $\lambda_c = \lambda (1 + \chi)$. 
Panels (b) and (d) depict the associated rectification ratios $R$.
Parameters are the same as in figure   \ref{fig:Current_vs_Coupling}.}
\label{fig:Current_vs_Asymmetry}%
\end{figure}

Figure \ref{fig:Current_vs_Asymmetry} examines the effect of asymmetry in the coupling strength on the diode effect. We test the effect at different coupling strengths $\lambda$ and for two temperature biases.
First, in  Fig. \ref{fig:Current_vs_Asymmetry} (a) and (c) we display the heat current itself in the forward direction using the BMR-QME and the EFF-SB methods.
Notably, the current is not maximal when the coupling is symmetric, 
rather, the current reaches an extremum when the qubit couples more strongly to the cold bath.
Fig. \ref{fig:Current_vs_Asymmetry}(b) and (d) display the corresponding rectification ratios  using the BMR-QME and the effective SB models. 
At stronger coupling the rectification ratio increases. However, this increase is rather minor.
Identifying strategies for significantly enhancing the thermal diode effect in 
impurity models remain a challenge.



\section{Summary}
\label{sec-sum}

We presented the reaction coordinate framework, RC-QME
for studies of quantum heat transport in nanoscale systems
beyond the weak system-bath coupling limit.
In this method, degrees of freedom of the baths, which are strongly coupled to the system are extracted and explicitly included into the system. After performing additional, physically-motivated transformations and truncation, 
transport characteristics of the enlarged system are computed using a perturbative, Markovian QME method. The main results of our work are:

(i) On the fundamental side, by continuing with  physically-motivated transformations of 
the RC Hamiltonian 
we ended up with a spin-boson model, albeit with renormalized parameters. The effective SB model
elegantly reveals the impact of strong system-bath coupling on spin dynamics and
heat transport behavior, by manifesting competing effects: When increasing the system-bath
coupling energy the effective spin splitting is strongly reduced. On the other hand,
the coupling of the effective spin to the baths increased approximately linearly.
One could generalize the principles behind the effective SB model, 
construct other effective few-level models with energy parameters renormalized 
to absorb strong coupling effects, and describe the operation of more complex thermal machines.   

(ii) On the application side,
by comparing our method to another scheme that utilizes the polaron transformation, we can safely conclude that our method captures system-bath coupling effects beyond low order perturbation theory treatments.
Specifically, the RC-QME method reproduced the turnover behavior of the heat current with $\lambda$, 
the system-bath interaction parameter, 
in a quantitative agreement with the polaron-NEGF method.
Furthermore, we demonstrated that the RC-QME method took into account off-resonant phonon transmission processes. 
This was manifested in the behavior of the heat current with temperature: 
The current showed multiple peaks representing resonant, then enhanced off-resonant transport 
once the temperature matched the peak of the bath density of states (spectral function).
We further looked at the diode behavior of the asymmetric SB model and 
showed that strong coupling effects could enhance
the rectification behavior, albeit to a small extent.

The construction of an effective SB model is one of the central results of this work, allowing us to 
rationalize strong coupling effects, as well as perform calculations at minimal computational cost.
Once reaching the effective SB model, quantum dynamics and transport behavior 
can be evaluated with a perturbative QME method, as we did here. 
Moreover, one can further employ the effective SB model in conjunction 
with more accurate techniques developed for the SB model,
such as the NIBA \cite{NIBA11} and Majorana-fermion NEGF methods \cite{NJP17,PT-NEGF17}. 
Even more so, once reaching the effective-SB model one could continue iteratively with the RC transformation, to extract 
additional modes into the system (transforming the spectral function numerically). 
As such, this technique is not limited to spectral functions that are peaked around a central mode.
Our method differs from chain mapping techniques such as TEPODA 
\cite{Plenio1,Plenio2,Plenio3}, 
since we are able to study the strong coupling regime semi-analytically, 
gaining physical intuition and minimizing computational costs.

The RC-QME method described in this paper should be extended in
two ways: First, one should test whether it is still accurate in models where the spectral density function of the SSB is not peaked at a single frequency as in the Brownian oscillator model, but rather resembles e.g. an Ohmic model. In such cases, one should find how many oscillators should be extracted from the baths to capture quantitatively strong coupling effects, and how does the reaction coordinate method correspond to the polaron model.
Second, the RC method and the effective Hamiltonian principle
can be feasibly generalized to describe the coupling of multi-level quantum systems to multiple reservoirs. As such, it is suitable to study the operation of strongly-coupled 
quantum heat machines, our objective in future studies.

\begin{acknowledgments}
The authors gratefully acknowledge 
Ahsan Nazir and Zachary Blunden-Codd for valuable discussions and their critical comments on the manuscript. The authors thank
Junjie Liu for fruitful discussions and for his guidance through the PT-NEGF method and code, which he had generously shared with the authors.
DS acknowledges funding from the Natural Sciences and 
Engineering Research Council (NSERC) of Canada Discovery Grant and the Canada Research Chairs Program. 
\end{acknowledgments}

\appendix
\renewcommand{\theequation}{A\arabic{equation}}
\setcounter{equation}{0}  
\section{Reaction coordinate mapping}\label{app:1}

The reaction coordinate mapping would be incomplete without addressing how the SSB and the 
RC spectral density functions are related to each other. 
We find this correspondence by studying the dynamics of the system in the two representations.
To generalize the discussion, we consider a coordinate $q$ moving in a confining potential $U(q)$.
We will make use of classical equations of motions   
but an analogous treatment can be done quantum mechanically. 
This can be done because the spectral density function only contains information about 
the harmonic baths and their interaction with the system, which is linear in the oscillators' displacements.
Our discussion here follows Ref. \cite{NazirPRA14}. It is included here for completeness, demonstrating its generalization for the case of a system coupled to two heat baths.

Starting from the SSB Hamiltonian Eq. (\ref{eq:HSB}), the classical analogue of the SSB model takes the form
\bea
H_q &=& \frac{P_q^2}{2} + U(q) + q\sum_{k,i} \phi_{k,i} x_{k,i} + q^2\sum_{k,i} \frac{\phi_{k,i}^2}{2\nu_{k,i}^2} 
\nonumber \\ 
&+& \frac{1}{2}\sum_{k,i} (p_{k,i}^2 + \nu_{k,i}^2 x_{k,i}^2). 
\label{HSB_q}
\eea
Here, we introduce the compact form $\phi_{k,i} \equiv \sqrt{2\nu_{k,i}}f_{k,i}$. 
Also, we identify the position and momenta of the environment as:
\bea
    x_{k,i} = \sqrt{\frac{1}{2 \nu_{k,i}}} (c_{k,i}^{\dagger} + c_{k,i})  
\eea
and
\bea
    p_{k,i} = i \sqrt{\frac{\nu_{k,i}}{2}} (c_{k,i}^{\dagger} - c_{k,i}).
\eea
Using this Hamiltonian, we obtain the classical equations of motion for the 
coordinate $q$ and each bath mode $x_{k,i}$:
\bea
&\ddot{q}(t)& = -U'(q) - \sum_{k,i} {\phi}_{k,i} x_{k,i}(t) - q(t)\sum_{k,i} \frac{{\phi}^2_{k,i}}{\nu_{k,i}^2}, 
    \\
    &\ddot{x}_{k,i}(t)& = -q(t){\phi}_{k,i} - \nu_{k,i}^2 x_{k,i}(t).
\eea
Here, $U'(q)$ corresponds to a derivative with respect to the coordinate $q$.
Working in Fourier space, where observables take the form $\check q(z) = \int_{-\infty}^{\infty} dt e^{-izt} q(t)$, it is possible to eliminate the environment variables from the equations of motion. 
Doing so, we obtain an algebraic equation $K(z) \check q(z) = -\check{U'}(z)$. 
Here, the Fourier space kernel $K(z)$ is defined as:
\bea
K(z) &=& -z^2\left( 1 + \sum_{k,i} \frac{{\phi}_{k,i}^2}{\nu_{k,i}^2(\nu_{k,i}^2 - z^2)} \right) 
    \nonumber \\
    &=& -z^2\left(1 + 2 \sum_i\int_{0}^{\infty}  d\nu \frac{J_{SSB,i}(\nu)}{\nu(\nu^2 - z^2)}  \right)
\nonumber\\
  &=& -z^2\left(1 +  \sum_i\int_{-\infty}^{\infty}  d\nu \frac{J_{SSB,i}(\nu)}{\nu(\nu^2 - z^2)}  \right).
\eea
The integral can be solved by the residue theorem making use of the simple pole 
at $\nu=z$. thus, the kernel takes the form:
\bea
K(z) = -z^2 + \pi i \sum_i J_{SSB,i}(z).
\eea
At this stage, we repeat the same procedure after the RC transformation, and identify the kernel 
$K(z)$ in terms of the RC spectral densities. 
Since the mapping is exact, the two kernel functions, before and after the normal mode 
transformation, should be identical. Following a similar procedure as before, we replace the Hamiltonian
 Eq. (\ref{eq:HRC}) by a classical coordinate moving in a potential $U(q)$. This yields 
\bea
    H_q &=& \frac{P_q^2}{2} + U(q) + q\sum_i \kappa_i x_i + q^2\sum_i \frac{\kappa_i^2}{2\Omega_i^2}
+ \frac{1}{2}\sum_i (p_i^2 + \Omega_i^2 x_i^2) 
\nonumber \\
&+& \sum_i x_i \sum_k {\xi}_{k,i} X_{k,i}
    + \frac{1}{2}\sum_{k,i}(P_{k,i}^2 + \omega_{k,i}^2 X_{k,i}^2)
   + \frac{1}{2}\sum_{k,i}\frac{\xi_{k,i}^2x_i^2}{\omega_{k,i}^2},
\nonumber\\
\label{eq:AHq}
\eea
where we define scaled coefficients $\kappa_i = \sqrt{2\Omega_i}\lambda_i$ 
and ${\xi}_{k,i} = 2g_{k,i}\sqrt{\Omega_i \omega_{k,i}}$, while $x_i$, $p_i$ and $X_{k,i}$, $P_{k,i}$ are positions and momenta of the RC and the environments, respectively. 
The Hamiltonian can be compactly written as 
\bea
H_q=H_s(q,x_i) +  \frac{1}{2}\sum_kP_{k,i}^2
+ \frac{1}{2}\sum_k \omega_{k,i}^2\left(X_{k,i}+x_i\frac{\xi_{k,i}}{\omega_{k,i}^2} \right)^2.
\nonumber\\
\eea 
$H_s(q,x_i)$ includes the terms from the first line of Eq. (\ref{eq:AHq}).
This Hamiltonian leads to equations of motion of the form
\bea
 &\ddot{q}(t)& = -U'(q) - \sum_i \kappa_i x_i(t) - q(t)\sum_i \frac{\kappa_i^2}{\Omega_i^2},
    \\
    &\ddot{x}_i(t)& = -q(t)\kappa_i - \left(\Omega_i^2 + \sum_k \frac{\xi_{k,i}^2}{\omega_{k,i}^2} \right)x_i(t) - \sum_k {\xi}_{k,i} X_{k,i}(t),
    \nonumber\\
    \\
    &\ddot{X}_{k,i}(t)& = -{\xi}_{k,i} x_{i}(t) - \omega_{k,i}^2 X_{k,i}(t). 
\eea
Working in Fourier space, it is possible to eliminate both the RC and environment variables from the equation of motion such that the kernel takes the form: 
\bea
K(z) = -z^2 + \sum_i \frac{\kappa_i^2}{\Omega_i^2} \frac{\mathcal{L}_i(z)}{\Omega^2_i + \mathcal{L}_i(z)};
    \label{eq:Kernel}
\eea
where, using the definition of the RC spectral density: 
\bea
    \mathcal{L}_i(z) &=& -z^2 \left( 1 + \sum_k \frac{{\xi}_{k,i}^2}{\omega_{k,i}(\omega_{k,i}^2 - z^2)} \right)
    \nonumber \\
    &=& -z^2 \left(1 + 4\Omega_i \int_0^{\infty} d\omega \frac{ J_{RC,i}(\omega)}{\omega(\omega^2 - z^2)} \right).
    \label{eq:L(z)}
\eea
If we assume that the RC spectral density is Ohmic, 
$J_{RC,i} =\gamma_i \omega e^{-|\omega|/\Lambda_i}$, 
where $\Lambda_i$ is a high frequency cut-off, then $\mathcal{L}_i(z) = -z^2 + 2i\pi z\Omega_i\gamma_i$.
By plugging this expression into Eq. (\ref{eq:Kernel}), writing $z=\omega - i\epsilon$, we 
equate the two expressions for the kernel $K(z)$. This leads to the expression,
\bea
J_{SSB,i}(\omega) &=& \frac{1}{\pi} \lim_{\epsilon \to 0^+} \Im \frac{\kappa_i^2}{\Omega_i^2} 
\frac{-(\omega-i\epsilon)^2 + 2i\pi\Omega_i \gamma_i (\omega-i\epsilon) }{\Omega^2_i - (\omega-i\epsilon)^2 + 2i\pi \Omega_i \gamma_i (\omega-i\epsilon)},
    \nonumber\\
    &=& \frac{4\gamma_i\omega \Omega_i^2\lambda_i^2}{(\Omega_i^2-\omega^2)^2 + (2\pi\gamma_i\Omega_i\omega)^2},
\eea
which is the associated spectral density function for the ``standard" spin-boson model.

\renewcommand{\theequation}{B\arabic{equation}}
\setcounter{equation}{0}  
\section{Rotated Reaction Coordinate Model}\label{app:2}

In this Appendix, we perform a rotation transformation on the RC Hamiltonian, Eq. (\ref{eq:HRC}), and absorb the quadratic term. This transformation is exact.
%
%
There are two terms, one for each RC, which are quadratic in the displacement of the reaction coordinate in Eq. (\ref{eq:HRC}),
$\sum_{i}(a_i^{\dagger} + a_i)^2\sum_k\frac{g^2_{k,i}}{\omega_{k,i}}$.
These terms can be absorbed 
by renormalizing the corresponding reaction coordinates.
To do this, we define a dimensionless parameter $\eta_i$ such that
\bea    
\eta_i\Omega_i &\equiv& \sum_k\frac{g^2_{k,i}}{\omega_{k,i}}
\nonumber\\
&=&\int_0^{\infty}  \frac{J_{RC,i}(\omega)}{\omega}d\omega = \gamma_i \Lambda_i.
\label{eq:eta}
\eea
The last equality holds assuming Ohmic spectral functions for the residual baths.
We now perform a rotation transformation of the RC oscillators, 
\bea
\Tilde{a}_i &=& a_i\cosh{r_i} + a_i^{\dagger}\sinh{r_i},
\nonumber\\
\Tilde{a}_i^{\dagger} &=& a_i^{\dagger}\cosh{r_i} + a_i\sinh{r_i},
\eea
where
\bea
e^{r_i} \equiv \sqrt{\frac{\Tilde \Omega_i}{\Omega_i}}; \,\,\,\,\,\,\,
    \Tilde \Omega_i = \Omega_i\sqrt{1 + 4\eta_i}.
\eea
Applying this transformation on Eq. (\ref{eq:HRC}) we get (what we refer to as) the rotated RC Hamiltonian,
\bea
\tilde H_{RC} &=& \frac{\epsilon}{2}\sigma_z + \frac{\Delta}{2}\sigma_x 
+\sum_i \tilde \Omega_i \tilde a_i^{\dagger}\tilde a_i
+\sigma_z \sum_{i}  \tilde \lambda_i (\tilde a_{i}^{\dagger} + \tilde a_i)
\nonumber\\
&+&   \sum_{k,i}\omega_{k,i}b_{k,i}^{\dagger}b_{k,i} 
 + \sum_{i} (\tilde{a}_i^{\dagger}+ \tilde{a}_i)\sum_k \tilde g_{k,i} (b_{k,i}^{\dagger}+b_{k,i}),
\nonumber\\
\label{eq:HRCn_B}
\eea
where
\bea
\Tilde{\lambda}_i&=& \frac{\lambda_i}{(1 + 4\eta_i)^{1/4}},
\nonumber\\
\Tilde{\Omega}_i &=& \Omega_i\sqrt{1 + 4\eta_i}, 
\nonumber\\
\tilde g_{k,i} &=& \frac{g_{k,i}} {(1 + 4\eta_i)^{1/4}} \,\,\,\,\,\, {\rm or } \,\,\,\,\,\,
\tilde \gamma_i =  \frac{\gamma_i} {\sqrt{1 + 4\eta_i}},
\eea
with $ \eta_i = \frac{\gamma_i\Lambda_i}{\Omega_i}$.
The Hamiltonian in Eq. (\ref{eq:HRCn_B}) 
includes a spin coupled to two ``primary" harmonic oscillators, each coupled to secondary harmonic baths. 
We select the spectral functions of the residual baths in the rotated-RC basis to be Ohmic,
\bea
\Tilde{J}_{RC,i}(\omega) = \tilde \gamma_{i} \omega e^{-\omega/\Lambda_i},
\label{eq:JRCn_B}
\eea
and find after a long algebra that it reduces to 
%
\bea
J_{SSB,i}(\omega) 
&=&
\frac{4\tilde{\gamma}_i\omega \tilde{\Omega}_i^2\tilde{\lambda}_i^2}{(\tilde{\Omega}_i^2-\omega^2 - 4\tilde{\Omega}_i \int_0^\infty \frac{\tilde{J}_{RC,i}d\omega}{\omega})^2 + (2\pi\tilde{\gamma}_i\tilde{\Omega}_i\omega)^2},
\nonumber \\
&=& \frac{4\gamma_i\omega \Omega_i^2\lambda_i^2}{(\Omega_i^2-\omega^2)^2 + (2\pi\gamma_i\Omega_i\omega)^2}.
\eea
%
%
The relationship between the SSB model and the rotated-RC model is a useful result of this work as it is an exact mapping, however, it comes with a warning. 
The behavior of the rotated-RC Hamiltonian cannot be studied with a Markovian QME method, as we discuss in Sec. \ref{Sec:B}. 
Other techniques, which take into account environmental memory and strong correlation effects should be used for an accurate treatment of this model.   

%


\renewcommand{\theequation}{C\arabic{equation}}
\setcounter{equation}{0}  
\section{Suppression of the effective tunneling splitting with $\lambda$} \label{app:3}

In this Appendix, we provide a deeper understanding of the suppression of $\Delta_{eff}(\lambda)$ with $\lambda$, the system-bath coupling energy, as displayed in Fig. \ref{effTLS}. 
Towards this goal, we study a simpler, but closely related model to that of the RC Hamiltonian (\ref{eq:HESm}). It showcases an effective two-level behavior in the energy basis. 

In the local basis, the model in question consists a qubit with splitting $\Delta$  (Pauli operators $\sigma_{x,y,z}^S$), coupled via $\lambda$ to a single harmonic oscillator with frequency $\Omega$, which is truncated at the second level ($M=2$). This two-state system (Pauli operators $\sigma^{RC}_{x,y,z}$ and identity operator $I^{RC}$) in turn is coupled to a heat bath held at a temperature $T$. Note, in order for this model to correspond to our simulations, we assume that $\Omega \gg \Delta, T$. Our system Hamiltonian is
\bea
H_{S} = \frac{\Delta}{2}\sigma_z^S +
\Omega\left( I^{RC} - \frac{1}{2}\sigma_z^{RC} \right)
+ \lambda \sigma_x^{S}\sigma_x^{RC}.
\label{eq:Hsys_Appen}
\eea
Both the qubit and the truncated oscillator have two possible states: The qubit can be in the states $\ket{\uparrow}$ or $\ket{\downarrow}$ while the oscillator can be in the occupation states $\ket{0}$ or $\ket{1}$. Therefore, the system can be in one of the four composite states: $\ket{\uparrow,0}$, $\ket{\uparrow,1}$, $\ket{\downarrow,0}$, and $\ket{\downarrow,1}$. In this local basis, the Hamiltonian is given by
\bea
H_S = \frac{1}{2}\begin{pmatrix} \Delta + \Omega & 0 & 0 & 2\lambda \\ 0 & \Delta + 3\Omega & 2\lambda & 0 \\ 0 & 2\lambda & -\Delta + \Omega & 0 \\ 2\lambda & 0 & 0 & -\Delta + 3\Omega \\  \end{pmatrix}.
\label{eq:matrix}
\eea
We diagonalize the Hamiltonian and get the following eigenvalues, ranked from highest to lowest energy; recall that $\Omega\gg \Delta$, and we also assume that $\Omega>\lambda$, 
\bea
E_{D}
&=& \Omega + \frac{1}{2}\sqrt{\Delta^2 + 4\lambda^2 + 2\Delta\Omega + \Omega^2} \,\,\,\,\,\,\, 
\xrightarrow{\lambda \to 0} \,\,\,\,\,\,\,
E_{\ket{\uparrow,1}} 
\nonumber\\
E_C &=& \Omega + \frac{1}{2}\sqrt{\Delta^2 + 4\lambda^2 - 2\Delta\Omega + \Omega^2}\,\,\,\,\,\, 
\xrightarrow{\lambda \to 0} \,\,\,\,\,\,\,
E_{\ket{\downarrow,1}}
\nonumber\\
E_B &=& \Omega - \frac{1}{2}\sqrt{\Delta^2 + 4\lambda^2 - 2\Delta\Omega + \Omega^2}\,\,\,\,\,\, 
\xrightarrow{\lambda \to 0} \,\,\,\,\,\,\,
E_{\ket{\uparrow,0}}
\nonumber\\
E_{A} &=& \Omega - \frac{1}{2}\sqrt{\Delta^2 + 4\lambda^2 + 2\Delta\Omega + \Omega^2}
\,\,\,\,\,\, 
\xrightarrow{\lambda \to 0} \,\,\,\,\,\,\,
E_{\ket{\downarrow,0}}
\nonumber\\
\eea
As in our effective two-state model in the main text, we assume that the two highest energy states are thermally inaccessible since $\Omega \gg T$. Instead, we focus on the lowest-lying states with an energy gap $\Delta_{eff}(\lambda) = E_{B}-A_A$, 
%
\bea
\nonumber
\Delta_{eff}(\lambda) &=& \frac{1}{2}\sqrt{\Delta^2 + 4\lambda^2 + 2\Delta\Omega + \Omega^2}
\\ \nonumber
&-& \frac{1}{2}\sqrt{\Delta^2 + 4\lambda^2 - 2\Delta\Omega + \Omega^2},
\\ \nonumber
&=& \frac{\Omega}{2}\left(1 + \frac{\Delta^2 + 4\lambda^2 + 2\Delta\Omega}{\Omega^2}\right)^{1/2}
\\
&-& \frac{\Omega}{2}\left(1 + \frac{\Delta^2 + 4\lambda^2 - 2\Delta\Omega}{\Omega^2}\right)^{1/2}.
\label{eq:Deltaeff}
\eea
Expanding Eq. (\ref{eq:Deltaeff}) up to third order in the small parameter $\frac{\Delta^2 + 4\lambda^2 \pm 2\Delta \Omega}{\Omega^2}$ using the binomial formula, $(1+x)^{1/2} \approx 1 -\frac{1}{2}x + -\frac{1}{8} x^2 + \frac{1}{16}x^3$, for $|x| \ll 1$, we obtain an approximate low order expression for the effective two level system splitting,
\bea
\Delta_{eff}(\lambda) \approx \Delta\left(1 - \frac{2\lambda^2}{\Omega^2} 
+ \mathcal{O}\left(\frac{\lambda^4}{\Omega^4}\right) +
\mathcal{O}\left(\frac{\lambda^2\Delta^2}{\Omega^4}\right)
+\mathcal{O}\left(\frac{\Delta^4}{\Omega^4}\right)\right).
\nonumber\\
\eea
This expression demonstrates the suppression of the 
effective splitting at strong enough coupling---once $\lambda/\Omega$ cannot be neglected.
We can further approximate it by
\bea
\Delta_{eff}(\lambda) \approx \Delta e^{-2\lambda^2/\Omega^2}.
\eea
This function was used in the main text to fit $\Delta_{eff}(\lambda)$ as displayed in Fig. (\ref{effTLS})(a).

The discussion in this section was confined to the case of a single RC, and truncating it most extremely. By coupling the spin to two RCs and repeating this calculation one would find
that $\Delta_{eff}({\vec \lambda})$ and $f({\vec \lambda})$ 
depend on the two RCs in a non-additive manner; 
${\vec \lambda}=(\lambda_1,\lambda_2)$.

\vspace{5mm}
\renewcommand{\theequation}{D\arabic{equation}}
\setcounter{equation}{0}  
\section{Additional simulations}\label{app:4}

\begin{figure}[h!]
    \centering
    {{\includegraphics[width=8.55cm] {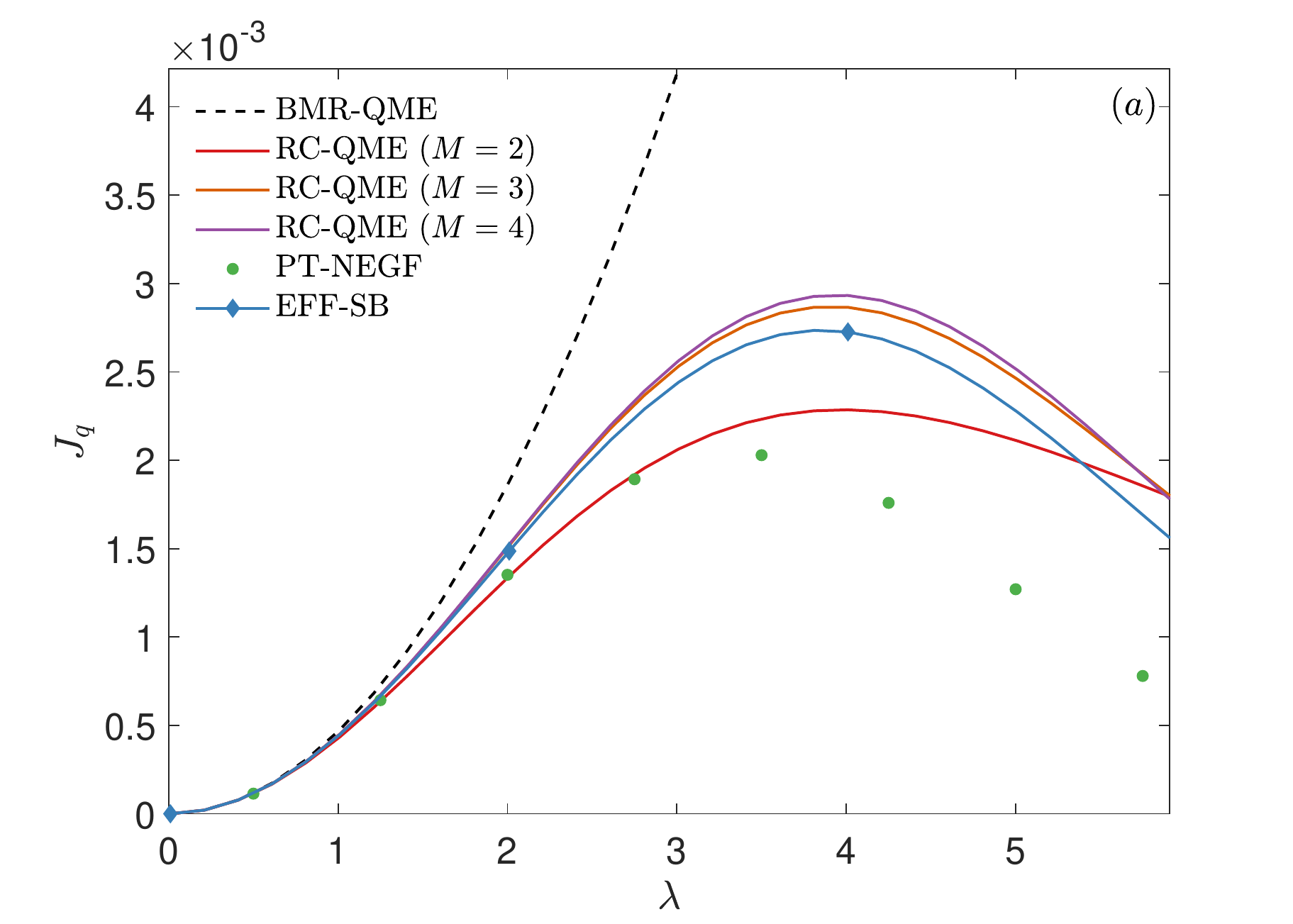}}} 
    \qquad
    {{\includegraphics[width=8.55cm]{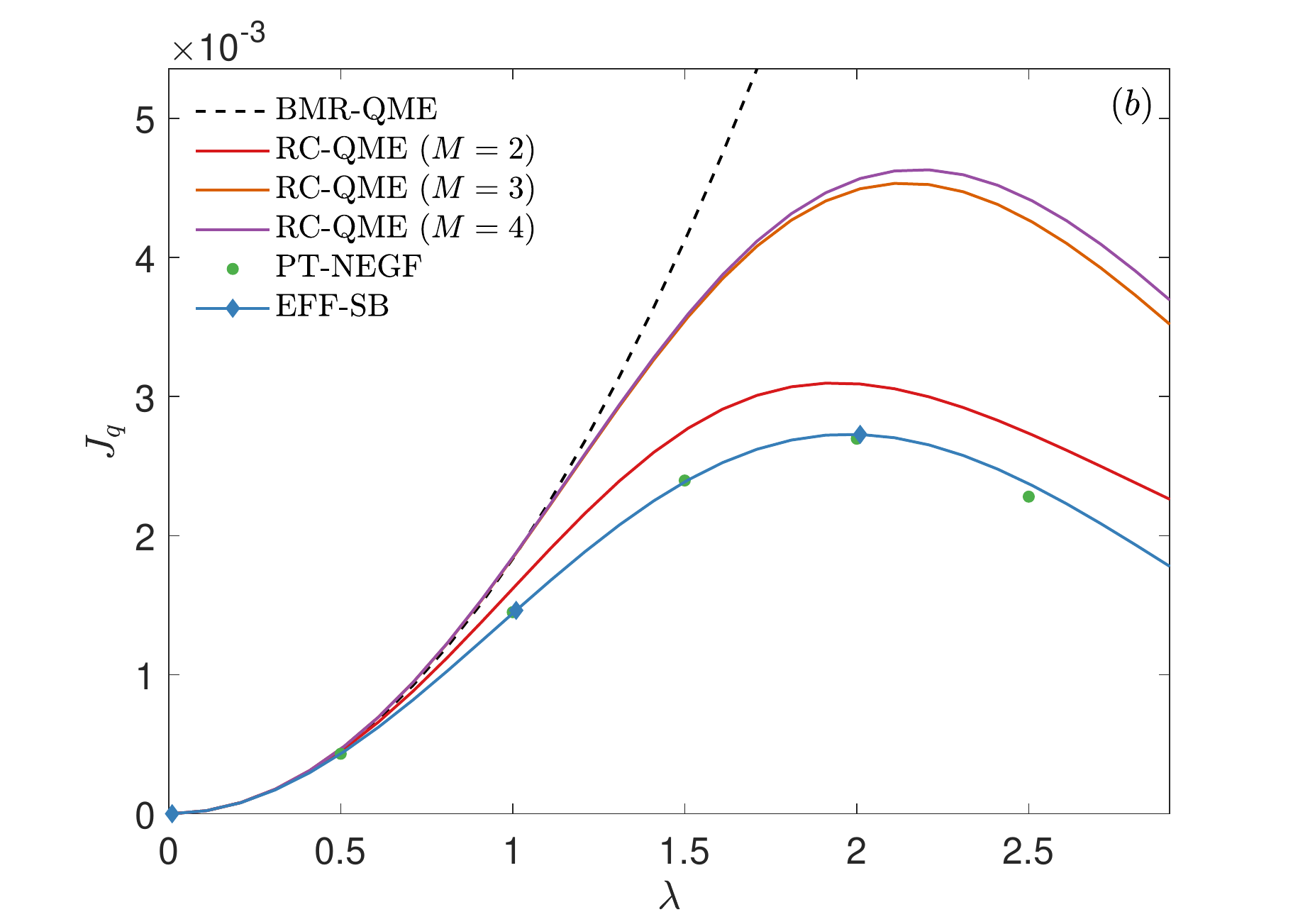}}}
    \caption{Steady state current as a function of  coupling strength $\lambda$. We compare simulations of the RC-QME ($M = 2-4$) to the BMR-QME (dashed), the PT-NEGF ($\circ$) and the EFF-SB ($\diamond$) methods.
    Parameters are $\epsilon = 0$, $\Delta = 1$, $\Lambda = 1000\pi\Delta$. The temperatures are $T_h = \Delta$, $T_c = 0.5\Delta$.
    (a) $\gamma = 0.005$, $\Omega = 10 \Delta$.
    (b) $\gamma = 0.005$,  $\Omega = 5 \Delta$.}       %
    \label{fig:Appendix_C}%
\end{figure}

In this Appendix we include additional simulations, displaying the behavior 
of the  steady state heat flux as a function of $\lambda=\lambda_i$.
These simulations, presented in Fig. \ref{fig:Appendix_C} demonstrate that general trends observed in Figs. \ref{fig:Current_vs_Coupling}-\ref{fig:Current_vs_lambda10}
hold even when we further reduced $\Omega$, the center of the spectral function in the SSB model.

As in the main text, we compare
the RC-QME results to BMR-QME, EFF-SB and the PT-NEGF methods. We assume symmetric reservoirs for simplicity and vary the RC frequency $\Omega$ in the two panels.
We find that the BMR-QME grows parabolically and diverges at large coupling in both panels while all strong-coupling techniques exhibit a turnover behaviour at roughly the same value.

Worthy of note, the width of the spectral function $J_{SSB}(\omega)$ centered around $\Omega$ in the SSB picture is controlled by the combination $\gamma\Omega$. Therefore, in panels (a) and (b) the associated spectral density function is broader relative to the one studied in the main text. 
This implies that there are more modes near $\Omega$ that are strongly coupled to the system, yet they are not extracted from the baths with a single reaction coordinate transformation. 
Therefore, treating the enlarged system using a perturbative quantum master equation approach becomes less accurate when the width parameter grows.
 This explains the large overshoot of the current.
 In order to accurately simulate the current in this regime, one would probably need to extract several modes (``chain mapping") from the reservoirs by iteratively performing the RC transformation.    


Remarkably, the EFF-SB treatment successfully captures the heat transport characteristics, compared to the PT-NEGF method, at a much cheaper computational cost, performing better than the $M$ level RC method. We discuss the superior performance of the EFF-SB method in Sec. \ref{sec:excep}.

\bibliographystyle{iopart-num}
\bibliography{RC12}

\begin{thebibliography}{10}
\bibitem{commentP}
   This comment is meant to explain seemingly odd values selected for
  parameters, $\Omega=28\Delta$ and $\gamma=0.0071/\pi$. Most importantly,
  these parameters satisfy $\Omega\gg \Delta$, beneficial for the effective
  spin--boson description, and $\gamma \ll1$, allowing us to use the RC--QME
  weak coupling method. Specific values for $\Omega$ and $\gamma$ were
  originally chosen based on parameters in the rotated picture, described in
  Appendix B. However, since we do not show results in the rotated basis, this
  reference point is irrelevant and inconsequential.
\end{thebibliography}
\begin{thebibliography}{99}
\bibitem{commentP}
This comment is meant to explain seemingly odd values selected for parameters,
$\Omega=28\Delta$  and $\gamma=0.0071/\pi$.
Most importantly, these parameters satisfy
$\Omega\gg \Delta$, beneficial for the effective spin-boson description,
and $\gamma \ll1$, allowing us to use the RC-QME weak coupling method.
Specific values for $\Omega$ and $\gamma$ were
originally chosen based on parameters in the rotated picture, described in
Appendix D. However, since
we do not show results in the rotated basis, this reference point is irrelevant
and inconsequential.
\end{thebibliography}

\end{document}

same transformations hold for dynamics - any message here?

Check convergence with M

secular-nonsecular?

Can we use the SB model with an Ohmic function, and still perform the RC numerically. Do we capture strong coupling?
'
Confusion: The Debye-Drude function

Benchmark: Junjie code, HEOM, polaron,...

Cooperative effects: Can we quantify them?  RC for one bath only?

Demonstrate enhanced/degraded performance at strong coupling
(heat current, QAR, diode)

Aug 28: Appendix A, sort factors of 2; Rectification at strong couplings? numerical procedure.

Sep 03: Analytic diagonalization, still need to sort factor 2 in Appendix. effective two state model can reveal max rectification
based on the HeffSB. Can we get lambda_m analytically?
Relation to polaron model.
Cooperative effects hidden in f(lambda_L, lambda_R)

* Establish the effective 2state model:
T, asymmety
Why is the effective TLS model valid for a symmetric system and less so for the asymmetric system?

--------------------------------
Separate ideas/studies

3 qubit model - 
dynamics of the SB model
diode at strong coupling (Junjie, Bijay, polaron, RC,polaron,...)
QAR at strong coupling Refrigerators with the RC 

-------------
Discuss: RC for e-ph systems
With the effective mode description, we can converge taking large M, 
and just not doing the full Redfield calcualtion. 
We can also generalize this concept, truncate the Hamiltonian in the eigenenergy basis to include 
say 6 states (each HO possibly with 1 excitation)

The effective Hamiltonian  eff-SB 

standard SB model --> a better name?

Bijay method

Appendix A - without quadratic term, quantum treatment

Appendix B: Figure 4 with other paramters

Dynamics

Beyond the Brownian oscillator model

refs and discussion: TEPODA,